\documentclass[11pt, a4paper]{article}
\pdfoutput=1

\usepackage{amsmath,amssymb,amsfonts,mathtools}
\usepackage{multirow}
\usepackage[utf8]{inputenc}
\usepackage{bm}
\usepackage{cite}
\usepackage{pifont}

\usepackage{physics}
\usepackage{slashed}
\usepackage{braket}
\usepackage{enumitem}
\usepackage{xargs}
\usepackage[footnotesize]{caption}
\usepackage{color}

\usepackage{ifpdf}
\ifpdf
  \usepackage{graphicx, hyperref, xcolor}     %   usepackage without driver option
\else     % For (p)LaTeX + dvipdfmx
  \usepackage[dvipdfmx]{graphicx, hyperref, xcolor}     %   usepackage with driver option
 \fi
\usepackage{subcaption}

\definecolor{rossoferrari}{HTML}{D9073D}
\definecolor{mediumblue}{HTML}{0000CD}
\definecolor{forestgreen}{HTML}{228B22}
\definecolor{desy_blue}{HTML}{009EE2}
\definecolor{desy_orange}{HTML}{FD8800}
\definecolor{light_pink}{rgb}{1,0.4,0.4}
\definecolor{light_blue}{rgb}{0.284602,0.317763,0.963947}
\hypersetup{% hyperref option list
setpagesize=false,
bookmarksnumbered=true,%
bookmarksopen=true,%
colorlinks=true,%
linkcolor=light_blue,
urlcolor=rossoferrari,
citecolor=rossoferrari,
linktocpage=false,
}

\usepackage[height=8.85in,width=6.8in]{geometry}

\usepackage{fourier}
\makeatletter
\@addtoreset{equation}{section}

\makeatother

\begin{document}
\begin{titlepage}

\begin{center}

\hfill KEK-TH-2431

\vskip 1.in

{
%\Huge \bfseries
%Axion induced chiral asymmetry\\[.5em]
%Time-dependent $\theta$ angle in QED
{\fontsize{24pt}{26pt}\selectfont \bfseries
%Understanding anomalous particle production\\ 
%in axion electrodynamics\\
%QED\\
Understanding anomalous particle production\\in massless QED via time-varying $\theta$ angle\\
}
%Understanding anomalous production of \\[.3em]
%fermions in axion QED
}

\vskip .8in

{\Large
Yu~Hamada$^{1}$, Ryuichiro~Kitano$^{1,2}$, Ryutaro~Matsudo$^{1}$, Kyohei~Mukaida$^{1,2}$
}

\vskip .6in
{\small
\begin{tabular}{ll}
$1$&\!\!\!\!\!\! \emph{KEK Theory Center, Tsukuba 305-0801, Japan}\\[.3em]
$2$&\!\!\!\!\!\! \emph{Graduate University for Advanced Studies (Sokendai), Tsukuba 305-0801, Japan}\\[.3em]
\end{tabular}
}
\end{center}
\vskip .6in

\begin{abstract}
\noindent 
The Maxwell equations imply that, under the background of non-zero
$\bm{B}$, varying $\theta$ term produces $\bm{E} \cdot \bm{B}$. An
interesting example is the Witten effect where a magnetic monopole
becomes a dyon which, however, should disappear in the exact massless
limit of the fermion. Underlying mechanism of this phenomenon has been
understood by Callan by the presence of an effective axion-like degree
of freedom around the monopole, which is roughly the phase of the
fermions. The configuration of this axion cancels the effect of the
$\theta$ term.
Now, the chiral anomaly implies that non-vanishing $\bm{E} \cdot
\bm{B}$ induces the chiral charge in the system. 
The question is whether the chiral charge is generated in the massless
limit when we take into account the axion-like degree of freedom in
the discussion.
The discussion is relevant for the mechanism of baryogenesis under the
background of time-dependent $\theta$.
We solve the system of the massless QED with time
dependent $\theta$ by reducing it to the two-dimensional QED.
We demonstrate the occurrence of chiral charge generation in the
background of static magnetic field for two cases: a magnetic monopole
and a uniform magnetic flux. For the monopole case, the chiral charge
comes out from the monopole while canceling the Witten effect. For the
case of the uniform flux, on the other hand, the effect of the
backreaction cannot be ignored, giving a more non-trivial time
dependence. We also discuss their implications on baryogenesis.
\end{abstract}

\end{titlepage}

\renewcommand{\thepage}{\arabic{page}}
\setcounter{page}{1}
%%%%%%%%%%%%%%%%%%%%%%%%%%%%%%%%%%%%%%%%%%%%%%%%%%

%%%%%%%%%%%%%%%%%%%%%%%%%%%%%%%%%%%%%%%%%%%%%%%%%%
\tableofcontents
\pagebreak
\renewcommand{\thepage}{\arabic{page}}
\renewcommand{\thefootnote}{$\natural$\arabic{footnote}}
\setcounter{footnote}{0}
%%%%%%%%%%%%%%%%%%%%%%%%%%%%%%%%%%%%%%%%%%%%%%%%%%

%%%%%%%%%%%%%%%%%%%%%%%%%%%%%%%%%%%%%%%%%%%%%%%%%%
\section{Introduction}
\label{sec:intro}

When the $\theta$ parameter is promoted to a field, \textit{i.e.}, to the
axion~\cite{Weinberg:1977ma,Wilczek:1977pj,Peccei:1977hh,Peccei:1977ur}, various counter intuitive phenomena occur in QED. In the
presence of a magnetic field, one can see from the Maxwell equations
that the time dependence of the axion induces an electric current
along the magnetic field. 
This flow of charge can explain the Witten effect~\cite{Witten:1979ey} on the magnetic
monopole. When we start with a monopole at $\theta = 0$ and slowly
change the $\theta$ term to a non-zero value, the electric charge
flows into (or out from) the monopole so that the monopole obtains a
non-vanishing electric charge at $\theta \neq 0$.
The result is similar in the case of a uniform static magnetic field
background. The flow of the charge is generated by the time
dependence of $\theta$, which results in the electric field along the
magnetic field.

When the system couples to fermions, the situation gets even more
mysterious. Through the chiral anomaly~\cite{Adler:1969gk,Bell:1969ts}, the operator $\bm{E} \cdot
\bm{B}$ is equivalent to the divergence of the chiral current $\partial \cdot J_5$. This
would imply the simultaneous generation of the chiral charge together with $\bm{E} \cdot \bm{B}$ by the time dependence
of $\theta$.
The mechanism of the chiral charge generation in the presence of the
magnetic field is a generalized version of the ``chiral magnetic
effect''~\cite{Vilenkin:1980fu,Alekseev:1998ds,Kharzeev:2007tn,Kharzeev:2007jp,Fukushima:2008xe,Kharzeev:2009pj,Son:2012wh,Zyuzin:2012tv} in QED.
%and/or QCD system, and can be directly understood from the Wess-Zumino-Witten (WZW) terms.

There is an important application of the phenomena in cosmology. 
By identifying the chiral charge here as the baryon number in the Standard Model, there have been extensive discussions~\cite{Bamba:2006km,Anber:2015yca,Jimenez:2017cdr,Domcke:2018eki,Domcke:2019mnd,Domcke:2020quw} of baryogenesis through the generation of $\bm{E} \cdot \bm{B}$ of $U(1)_Y$ by the slow rolling of the axion (like field) in the early Universe.\footnote{
See also Refs.~\cite{Joyce:1997uy,Fujita:2016igl,Kamada:2016eeb,Kamada:2016cnb} for baryogenesis through the decaying helical $U(1)_Y$ magnetic field.
%The baryon asymmetry is also generated through the decay of this helical $U(1)_Y$ magnetic field at later stages~\cite{Joyce:1997uy,Bamba:2006km,Anber:2015yca,Kamada:2016eeb,Kamada:2016cnb,Jimenez:2017cdr}.
%%
}
This mechanism provides an elegant realization of spontaneous baryogenesis~\cite{Cohen:1987vi,Cohen:1988kt,Cohen:1991iu}. The amount of baryon asymmetry in the Universe can be explained by this mechanism.
(See also Refs.~\cite{Turner:1987bw,Garretson:1992vt,Anber:2006xt,Caprini:2014mja,Adshead:2016iae,Caprini:2017vnn} for productions of primordial magnetic field from the rotating axion.)

However, the direct relation between $\partial \cdot J_5$ and $\bm{E} \cdot \bm{B}$ is true only in the case of the massless fermions which make the discussion somewhat complicated.\footnote{
  The effect of a fermion mass on the chiral asymmetry generation is discussed for instance in Refs.~\cite{Warringa:2012bq,Copinger:2018ftr,Domcke:2019qmm}.
}
%the anomaly relation is true only in the case of the massless fermions which make the discussion somewhat complicated. 
In the presence of the massless fermions, the generation of $\bm{E} \cdot \bm{B}$ and the generation of the chiral charge from that cannot be separately considered in general.
This is because the backreaction from the chiral charge production cannot be neglected in the Maxwell equations, which is represented as the induced current. 
%Also, since the anomaly is a quantum effect, one should consider quantization of fermions in a time-dependent background.
Furthermore, we should consider the simultaneous production of the chiral charge and $\bm{E} \cdot \bm{B}$ with quantum effects taken into account since the anomaly is a quantum effect.\footnote{
  In cosmological applications for instance, such backreaction has been mostly neglected except for Refs.~\cite{Domcke:2018eki,Domcke:2019mnd,Domcke:2019qmm,Domcke:2020quw,Gorbar:2021rlt,Gorbar:2021zlr,Fujita:2022fwc}.
  Yet, even in these studies, the simultaneous production of the chiral charge and $\bm{E} \cdot \bm{B}$ with quantum effects has not been addressed.
}

Indeed, in the presence of the massless fermions, the Witten effect is
absent since the theory cannot be $\theta$ dependent. This fact may
indicate that the chiral charge generation is absent or suppressed.
The absence of the Witten effect in the massless fermion limit has
been understood by the dynamics of the fermions in the s-wave state
around the monopole~\cite{Callan:1982ah,Callan:1982au,Rubakov:1982fp}. These fermions exhibit condensations near the
monopole and an axion like degree of freedom appears. 
This axion actually cancels the effects of the $\theta$ term so that the theory
is $\theta$ independent near the monopole. In the massless limit,
everywhere gets ``near the monopole'' and eliminates the $\theta$
dependence everywhere. The massless limit is in this way smoothly
connected from the massive case. Around the monopole, there is a
$\theta$ independent region with a radius of the order of the inverse
of the fermion mass, and the region gets infinitely large in the
massless limit~\cite{Yamagishi:1982wp}.
This discussion suggests that one should include an axion-like degree of freedom in
the dynamics when we consider the massless fermions.

In this paper, along the line of the discussion of the Witten effect
with fermions, we study the chiral charge generation by the time
dependent $\theta$ in the background of the static magnetic field.
We discuss both in the cases of the monopole and the uniform magnetic
fields with the simplification to reduce the system to two-dimensional
QED by considering only the spherically symmetric configurations for
the monopole or the lowest Landau level in the case of the uniform
magnetic field. This treatment is expected to fully reproduce the
effects of the anomaly as higher modes do not contribute to the
anomaly.
By the reduction to the two-dimensional system, one can use the
technique of the bosonization which makes it possible to describe the production of the chiral charge and $\bm{E} \cdot \bm{B}$ simultaneously with full quantum effects taken into account.
In both cases, we find analytic solutions to the equations of motion,
and the time dependence of the electric charge and chiral charge can be fully
understood.

\smallskip
\paragraph{\textit{Organization of this paper.---}}
This paper is organized as follows. 
Our starting point is the following massless QED with a time-dependent $\theta$ term:
\begin{equation}
  \label{eq:setup}
  S = \int \dd^4 x\, \qty[ 
    - \frac{1}{4} F_{\mu \nu} F^{\mu\nu} + \overline{\psi} i D_\mu \gamma^\mu \psi
    + \frac{g^2 \theta (t)}{16 \pi^2} F_{\mu\nu} \tilde F^{\mu\nu}
   ],
\end{equation}
where the covariant derivative is $D_\mu \equiv \partial_\mu + i g
A_\mu$, and the field strength and its dual are denoted as
$F_{\mu\nu}$ and $\tilde F^{\mu\nu} \equiv \epsilon^{\mu\nu\rho\sigma}
F_{\rho\sigma} / 2$ respectively with the total anti-symmetric tensor
being $\epsilon^{0123} = + 1$. The conservation of the $U(1)_A$
current, $J_5^\mu \equiv \overline{\psi} \gamma^\mu \gamma_5 \psi$, is
broken by the topological density via the chiral anomaly
\begin{equation}
  \partial_\mu J_5^\mu = - \frac{g^2}{8 \pi^2} F_{\mu\nu} \tilde F^{\mu\nu}.
\end{equation}
When there is a massless fermion, the theory does not depend on the ``$\theta$-parameter'' (\textit{i.e.}, constant $\theta$) because the $U(1)_A$ transformation cancels the dependence.
We keep the argument of $\theta$ explicitly to clarify that we promote $\theta$ to a dynamical quantity.
We start with the discussion of the monopole in Section~\ref{sec:monopole}.
We briefly sketch how to derive the effective two-dimensional system in the s-wave approximation.
Then, by bosonizing the two-dimensional fermions, we solve the dynamics under the time dependence of $\theta$. 
In Section~\ref{sec:flux}, we consider the case of the uniform magnetic field.
Again we derive the effective two-dimensional system restricting ourselves to the lowest Landau level.
We solve it under the time dependence of $\theta$ by bosonizing the two-dimensional fermions.
The last Section~\ref{sec:concl} is devoted to conclusions and discussion.
%We provide a couple of constraints that should be imposed if we would like to use this mechanism as the origin of the baryon asymmetry of the Universe.

\newpage
%%%%%%%%%%%%%%%%%%%%%%%%%%%%%%%%%%%%%%%%%%%%%%%%%%
\section{Monopole}
\label{sec:monopole}
In this section, we study the chiral asymmetry production in the monopole background. 
Naively, the absence of the Witten effect seems to imply that the effect of the change of $\theta$, especially the chiral asymmetry generation, is cancelled.
However, as we will show in the following, this is not the case.
We first bosonize the effective two-dimensional system for the s-wave.
Then we see how the chiral asymmetry is generated \textit{classically} contrary to the naive expectation by solving the equation of motion.

\subsection{Bosonization under s-wave approximation}

In this subsection, we temporally introduce the mass of the fermion, and we will take the massless limit $m \to 0$ in the end.
Let us consider a monopole sitting at the origin, whose vector potential can be written in a spherical coordinate as
\begin{equation}
    \overline{A} \equiv \overline{A}_\mu \dd x^\mu = \frac{1}{2 g} \qty( - 1 + \cos \theta ) \dd \varphi.
\end{equation}
Here a singularity of $\overline{A}$ at $\theta = \pi$ is compensated by a gauge redundancy of $\overline{A} \sim \overline{A} - \dd \varphi / g$.
One may readily integrate its magnetic field over a $2$-sphere wrapping the origin and confirm that it is indeed a monopole, \textit{i.e.}, 
$(g / 2 \pi) \int_{S^2} \overline{B} = - (g / 2 \pi) \int_{S^2} \dd \overline{A} = 1$.
In the monopole background, the partial wave expansion can be applied due to the spherical symmetry.

In the following, we only consider the s-wave because the higher partial waves of the fermion field are kept away from the monopole core.
The s-wave approximation is performed by substituting the spherically symmetric form of the gauge field
\begin{equation}
    A = A_\mu \dd x^\mu = A_t \qty( t,r ) \dd t + A_r \qty( t,r ) \dd r + \overline{A},
    \label{eq:swave_gauge}
\end{equation}
and the fermion field
\begin{equation}
    \psi = \frac{1}{ \sqrt{4 \pi} r } \chi_+ \otimes
    \begin{pmatrix}
        \psi_{L,s} \qty(t,r)  \\
        \psi_{R,s} \qty(t,r)
    \end{pmatrix}, \qquad
    \chi_+ \equiv \begin{pmatrix}
        \cos \frac{\theta}{2} \\ e^{ i\varphi } \sin \frac{\theta}{2}
    \end{pmatrix}.
    \label{eq:swave_fermion}
\end{equation}
Here the $L$ and $R$ subscripts in the fermion field indicate its chirality in the original four-dimensional field.
In this basis, the four-dimensional gamma matrices $\gamma^\mu$ are given as $\gamma^0=\sigma^1\otimes 1_{2\times 2}$ and $\gamma^i = i \sigma_2\otimes \sigma^i$.
Let us emphasize again that, as the $\varphi$-dependence of $A$ and $\psi$ is compensated by a gauge redundancy of $( \psi, A ) \sim ( e^{i \varphi} \psi, A - \dd \varphi / g)$, we can identify a segment of $0 \leqslant \varphi \leqslant 2\pi$ at $\theta = \pi$ with a point. Hence this is a consistent configuration in a spherical coordinate.
Also, this configuration represents \textit{spherically symmetric} in the sense that the spatial rotation can be compensated by a gauge transformation. 
See Appendix~\ref{sec:appendix_spherical}.

By inserting these s-wave configurations, we find the s-wave action as follows
\begin{align}
    S = \int \dd t \int_0^\infty \dd r  \,
    \qty[ 2\pi r^2 F_{tr}^2  
    + i \psi_{L,s}^\dag (D_t - D_r) \psi_{L,s} 
    + i \psi_{R,s}^\dag (D_t + D_r) \psi_{R,s}
    - m \qty( \psi_{R,s}^\dag \psi_{L,s} + \text{H.c.} ) - \frac{g \theta}{2\pi}F_{tr} ].
 \label{swave}
\end{align}
Now it is clear that the effective s-wave theory around the monopole is nothing but a two-dimensional theory defined on a half-line.
For later convenience, we introduce the two-dimensional Dirac fields as $\psi_{\bm{2}} \equiv (\psi_{L,s}, \psi_{R,s})^T$.
the action reduces to the usual form
\begin{align}
  S &= \int \dd t \int_0^\infty \dd r\,  
  \qty[
    2\pi r^2 F_{tr}^2  + \overline{\psi}_{\bm{2}} (i\gamma^\mu_{\bm{2}} D_\mu - m)\psi_{\bm{2}} - \frac{g \theta}{2\pi}F_{tr}
  ],
  \label{eq:swave-dirac}
\end{align}
where $\mu=t,r$ and the two-dimensional gamma matrices $\gamma^\mu_{\bm{2}}$ defined as
\begin{align}
  \gamma^t_{\bm{2}} = \sigma^1, \quad \gamma^r_{\bm{2}} = i\sigma^2, \quad \gamma_{\bm{2}}^5 = \gamma^t_{\bm{2}} \gamma^r_{\bm{2}} = - \sigma^3.
\end{align}
In two dimensions, $\psi_{L,s}$ and $\psi_{L,s}^\dag$ are ``left-mover'' corresponding to incoming wave, and $\psi_{R,s}$ and $\psi_{R,s}^\dag$ are ``right-mover'' corresponding to outgoing wave.
They should not be confused with the handedness in four dimensions.

As the effective s-wave action of \eqref{eq:swave-dirac} describes a two-dimensional theory, fermions can be bosonized, namely we have an equivalent bosonic theory in two dimensions.
The bosonized action is given by 
\begin{equation}
  S_b = \int \dd t \int_0^\infty \dd r\, \qty{
    2\pi r^2 F_{tr}^2 + \frac{g}{2\pi} \qty(\phi - \theta) F_{tr} + \frac1{8\pi} \qty[ \qty(\partial_t\phi)^2 - \qty(\partial_r\phi)^2 ] 
    - \mu m \qty( 1-\cos\phi )
  },
  \label{eq:bosonization_1}
\end{equation}
where $\mu$ is a renormalization scale and $\phi$ is the phase of the fermion bilinear operator
\begin{equation}
  \psi_{R,s}^\dag \psi_{L,s} = -\frac12 \mu e^{i\phi}.
\end{equation}
Note that we normalize $\phi$ so as to be $2\pi$ periodic.
%As a rule of thumb, we can rewrite the two-dimensional $U(1)_V$ current and the $U(1)_A$ current by using this bosonic field $\phi$ as follows
The two-dimensional $U(1)_V$ current and the $U(1)_A$ current carried by the fermions are written using this bosonic field $\phi$ as
\begin{align}
  J_{\bm 2}^0 &\equiv \overline{\psi}_{\bm 2} \gamma^t_{\bm 2} \psi_{\bm 2} = - \frac{1}{2 \pi} \partial_r \phi ,\qquad
  J_{\bm 2}^r \equiv \overline{\psi}_{\bm 2} \gamma^r_{\bm 2} \psi_{\bm 2} +J_{\bm 2,\mathrm{CS}}^0= \frac{1}{2 \pi}\partial_t \phi, \label{eq:current}\\
  J_{\bm{2}, 5}^0 &\equiv \overline{\psi}_{\bm 2} \gamma^t_{\bm 2} \gamma_{\bm{2}}^5 \psi_{\bm 2} = \frac1{2\pi}\partial_t \phi,\qquad
  J_{\bm{2}, 5}^r \equiv \overline{\psi}_{\bm 2} \gamma^r_{\bm 2} \gamma_{\bm{2}}^5 \psi_{\bm 2} = -\frac{1}{2\pi}\partial_r\phi. \label{eq:chiral_current}
\end{align}
At $r\neq 0$, these currents fulfill the relation of $\epsilon^{\mu\nu} J_{\bm{2}, \nu} = - J_{\bm{2},5}^\mu$ as indicated by the property of two-dimensional gamma matrices, \textit{i.e.}, $\epsilon^{\mu\nu}\gamma_{\bm{2},\nu} = - \gamma_{\bm{2}}^\mu \gamma^5_{\bm{2}}$.
These two-dimensional currents are related to the original four-dimensional currents as 
\begin{align}
  J^0 \equiv \overline{\psi} \gamma^0 \psi = \frac{J^0_{\bm 2}}{4 \pi r^2} + \frac\phi{2\pi}\delta^{(3)}(\bm{x}), \quad
  \bm{J} \equiv \overline{\psi} \bm{\gamma} \psi =  \frac{\hat{\bm{r}}}{4 \pi r^2} J^r_{\bm 2}, \quad
  J^0_5 \equiv \overline{\psi} \gamma^0 \gamma_5 \psi = \frac{J^0_{\bm{2},5}}{4 \pi r^2}, \quad
  \bm{J}_5 \equiv \overline{\psi} \bm{\gamma} \gamma_5 \psi = \frac{\hat{\bm{r}}}{4 \pi r^2} J_{\bm{2},5}^r.
  %\bar\psi\gamma^0\psi
  %&\frac1{4\pi r^2}\bar\psi_S\gamma^t_S\psi_S = \bar a\bar\sigma^0 a - \bar b\bar\sigma^0b = \bar\psi\gamma^0\psi,\quad
  %\frac{\hat r^j}{4\pi r^2}\bar\psi_S\gamma^r_S\psi_S =\bar a\bar\sigma^j a - \bar b\bar\sigma^jb = \bar\psi\gamma^j\psi.
\end{align}
where the second term of $J^0$ is the contribution from the monopole, $\phi \nabla\cdot \bm{B}$.
Due to the contribution from the Chern-Simons coupling, the total electric current is
\begin{align}
  g J_\text{el}^\mu = gJ^\mu + J_{\mathrm{CS}}^\mu %\notag\\
  \quad \text{with}~~~
  %&
  J_{\mathrm{CS}}^0 \equiv \frac1{8 \pi^2 r^2}g\partial_r\theta - \frac1{2\pi} g\theta |_{r=0} \delta^{(3)}(\bm{x}),
  \quad J_{\mathrm{CS}}^r \equiv -\frac1{4\pi r^2}\frac{g\partial_t\theta}{2\pi}.
\end{align}
The two contributions $J^\mu$ and $J^\mu_{\mathrm{CS}}$ are separately conserved.

One can further simplify the bosonized action \eqref{eq:bosonization_1} by solving the equation of motion for the gauge field.
By taking the variation of $A_t$ and $A_r$ we obtain
\begin{align}
  \partial_r(4\pi r^2 F_{tr}) + \frac{g}{2\pi}\partial_r\qty(\phi-\theta) &= 0, \\
  \partial_t(4\pi r^2 F_{tr}) + \frac{g}{2\pi}\partial_t\qty(\phi-\theta) &= 0.
\end{align}
%Hereafter we take the temporal gauge $A_t = 0$.
%By completing the square, we can immediately integrate out $A_r$, which leads to the Gauss law constraint at the same time
%\begin{equation}
%   \partial_t E_r =
%  \partial_t^2 A_r = -\frac{1}{4\pi r^2} \frac{g}{2 \pi} \partial_t \qty(\phi - \theta).
%  \label{eq:E}
%\end{equation}
%\crm{Redefining $E_r:=F_{tr}=\partial_tA_r$.}
%This equation can be integrated trivially. 
%The constant part of the integration can be absorbed by redefining the $\theta$ parameter.
The general solution of these equations is
\begin{align}
  F_{tr} = -\frac1{4\pi r^2}\frac{g}{2\pi}\qty(\phi-\theta + c),
  \label{eq:F}
\end{align}
where $c$ is an integration constant.
The value of $c$ is automatically fixed as $c = -\phi(r=0)+\theta(r=0)$ because otherwise $F_{tr}$ would diverge at the origin, and then the action would also diverge.
Now we eliminate $c$ by redefining $\theta$ parameter.
Then we finally obtain the action only with the boson field $\phi$ as\footnote{
In the action (\ref{eq:action_boson}), the $2\pi$-periodicity of $\phi$ seems to be violated.
However, the $2\pi$-periodicity is maintained in the following way.
When we shift $\phi$ by $2\pi$, the constant $c$ in Eq.~(\ref{eq:F}) is automatically changed to satisfy $\phi(r=0)-\theta(r=0)+c=0$, and thus the value of $F_{tr}$ is unchanged.
By substituting $F_{tr}$, we obtain the same action as Eq.~(\ref{eq:action_boson}), and then the $2\pi$ shift does not change anything.
Note that when the mass is zero, the shift of $\phi$ by any real number does not change the action in the same way.
}
\begin{equation}
  S_b = \int \dd t \int_0^\infty \dd r \qty{
  \frac1{8\pi} \qty[(\partial_t\phi)^2 - (\partial_r\phi)^2] - \mu m \qty(1-\cos\phi) - \frac{1}{2}\frac{1}{4 \pi r^2} \qty(g \frac{\phi - \theta}{2 \pi} )^2 }.
  \label{eq:action_boson}
\end{equation}
%\crm{Added the explanation of the boundary condition.}
In order for the last term to be finite, the following boundary condition has to be satisfied:
\begin{align}
  \phi(r=0) - \theta(r=0) = 0.
\end{align}
Due to this condition, the boundary contributions from $\phi$ and $\theta$ to the net electric charge are cancelled out.

\subsection{Production of chiral asymmetry from the monopole}

\paragraph{Classical solution.}
Let us analyze the theory classically in the massless limit.
The equation of motion is
\begin{equation}
    \partial_t^2\phi - \partial_r^2\phi + \frac{1}{4 \pi r^2} \frac{g^2}{\pi} \qty(\phi - \theta ) = 0,
\end{equation}
which fulfills the anomaly equation automatically as one can see from Eqs.~\eqref{eq:current}, \eqref{eq:chiral_current} and \eqref{eq:F}.
We consider the following time evolution of $\theta$ term:
\begin{equation}
  \label{eq:theta_t}
  \theta (t) = 
  \begin{cases}
    0 & \text{for} \quad t < 0, \\
    v_\theta t & \text{for} \quad 0 \leqslant t,
  \end{cases}
\end{equation}
where we have taken the initial value of $\theta$ to be zero without loss of generality since a constant $\theta$ can be rotated out for massless fermions.
We also assume that there are neither net chiral charge nor electric charge initially.
This implies the following initial conditions, which represents \textit{classical} vacuum, 
\begin{equation}
  \phi (t = 0, r) = 0 , \quad \partial_t \phi (t = 0, r) = 0.
\end{equation}
By defining $\tilde \phi \equiv \phi - \theta$, one may rewrite the equation of motion in the following convenient way:
\begin{equation}
  \partial_t^2 \tilde \phi - \partial_r^2 \tilde \phi + \frac{1}{4 \pi r^2} \frac{g^2}{\pi} \tilde \phi = 0,
\end{equation}
with the initial conditions being $\tilde \phi (t=0,r) = 0,\ \partial_t \tilde \phi (t=0,r) = - v_\theta$.

\begin{figure}[t]
\centering
\includegraphics[width=0.5\hsize]{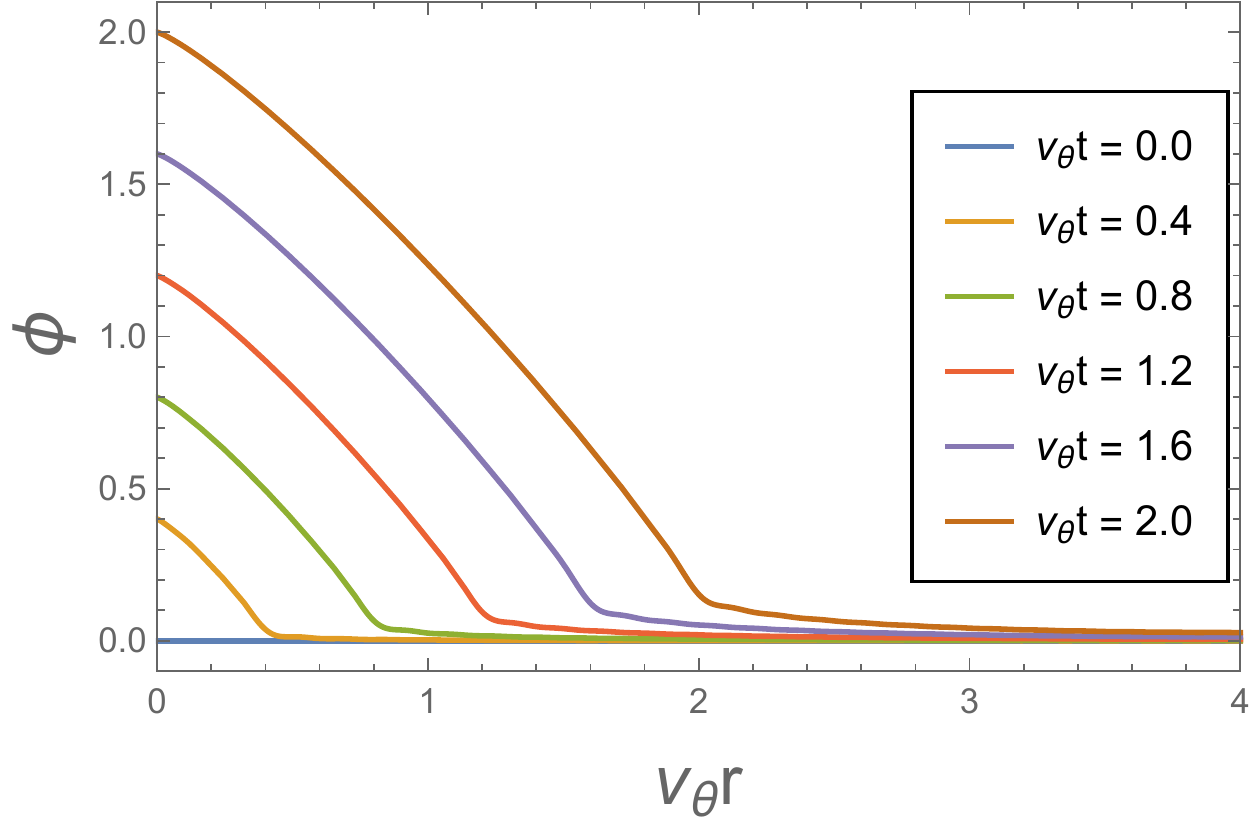}
\caption{The plot of $\phi$ for $v_\theta r_0=4.0$ and $\alpha=0.7$, where the sum is taken up to $n=50$.}
\label{fig:phi}
\end{figure}

For our numerical study, we approximate this theory on a half line with that on a segment of $0 \leqslant r \leqslant r_0$, whose boundary condition at $r = r_0$ is $\partial_r \tilde \phi (t,r_0) = 0$.
We expect that the behavior of the solution for $r,t \ll r_0$ is not affected by this boundary condition.
The solution is given as infinite series of
\begin{equation}
  \tilde \phi \qty(t,r) = \sum_{n=1}^\infty C_n \sin \qty(k_nt) j \qty(k_n r),
  \label{eq:solution_monopole}
\end{equation}
where $j(x) \equiv \sqrt{x} J_{\alpha}(x)$, $\alpha \equiv \sqrt{1/4 + g^2/(4 \pi^2)}$, $J_{\alpha}(x)$ is the Bessel function of the first kind, and $k_n \equiv z_n/r_0$ with the positive $n$-th zero of $j'(x)$ being $z_n$.
%\crm{I have replaced root by zero, matching the mathematica document.}
This is because $\{j(k_n r)\}$ forms a complete set of functions $f(r)$ with $f(0)=f'(r_0)=0$, which satisfies $\int_0^{r_0} \dd r  j(k_nr) j(k_mr) = N_n \delta_{nm}$ with $N_n$ being a normalization.
%\red{[YH: recursive definition for $z_n$? In a more simple way, $z_n$ can be defined as $z_n\equiv n \pi + (2\alpha+1)\pi/4$ because $\sqrt{x} J_\alpha(x)$ has an asymptotic form $\sqrt{x} J_\alpha(x) \sim \cos (x-(2\alpha+1)\pi/4)$ for $x\to  \infty$.]}
The expansion coefficient is obtained from
\begin{equation}
  N_n k_n C_n 
  = \int_0^{r_0} \dd r\,  \partial_t \tilde \phi \qty(t=0,r) j \qty(k_n r) 
  = - v_\theta \int_0^{r_0} \dd r\, j \qty(k_n r).
\end{equation}
We can numerically obtain the values of $N_n$, $k_n$ and hence $C_n$. See Fig~\ref{fig:phi}.

The qualitative feature of the solution can be approximated with that at the leading order in $g$
\begin{equation}
  \label{eq:monopole_sol_aprx}
  \phi(t,r) \simeq \begin{cases}
    v_\theta (t-r) &\text{for} \quad 0 < r < t,\\
    0      &\text{for} \quad t < r \ll r_0.
  \end{cases}
\end{equation}
This gives the evolution of the currents and the electric field as follows
\begin{equation}
    J^0 \simeq \frac\theta{2\pi}\delta^{(3)}(\bm{x}) + \frac{1}{4 \pi r^2} 
    \begin{cases}
      \frac{v_\theta}{2 \pi}\\
      0
    \end{cases}\!\!\!\!\!\!\!\!,
    \quad
    \bm{J} \simeq \frac{\hat{\bm{r}}}{4 \pi r^2} 
    \begin{cases}
      \frac{v_\theta}{2 \pi} \\
      0
    \end{cases}\!\!\!\!\!\!\!\!, \quad
    J^0_5 \simeq \frac{1}{4 \pi r^2} 
    \begin{cases}
      \frac{v_\theta}{2 \pi} \\
      0
    \end{cases}\!\!\!\!\!\!\!\!, \quad
    \bm{J}_5 \simeq \frac{\hat{\bm{r}}}{4 \pi r^2}
    \begin{cases}
      \frac{v_\theta}{2\pi} \\
      0
    \end{cases}\!\!\!\!\!\!\!\!, \quad
    E^r \simeq \frac{-g}{4 \pi r^2} \begin{cases}
      \frac{v_\theta r}{2 \pi} \\
      \frac{v_\theta t}{2 \pi}
    \end{cases}\!\!\!\!\!\!\!\!, 
    \label{eq:quantities}
\end{equation}
for $0 < r < t$ and $t < r \ll r_0$ respectively.
Note that the electric current from the Chern-Simons coupling is given by
\begin{equation}
  \bm{J}_\text{CS} = - \frac{\hat{\bm{r}}}{4 \pi r^2} \frac{g v_\theta}{2 \pi}
\end{equation}
for $t > 0$ and vanishing otherwise.

The obtained solution has a clear physical picture as follows.
As can be seen from the evolution of the phase~\eqref{eq:solution_monopole} and also the currents $\bm{J}$ and $\bm{J}_5$ in \eqref{eq:quantities}, the fermions are generated from the monopole and propagate at the speed of light.
This phenomenon of fermion number violation is similar to the Callan-Rubakov effect in the monopole-fermion scattering where the monopole becomes the source of anomalous production of the chiral charge~\cite{Callan:1982ah,Callan:1982au,Rubakov:1982fp}.
In the case of dynamical $\theta$,
interestingly, the fermions are emitted from the monopole so that they exactly cancel out the incoming electric current from the Chern-Simons coupling up to the radius of the light front, \textit{i.e.},
\begin{equation}
    g\bm{J}_\text{el} \simeq
    \frac{\hat{\bm{r}}}{4 \pi r^2} \begin{cases}
    0 &\text{for} \quad 0 < r < t, \\
    - \frac{g v_\theta}{2 \pi} &\text{for} \quad t < r \ll r_0.
\end{cases}
\end{equation}
We now see the absence of the Witten effect with massless fermions since the electric current never reaches the core of the monopole.
This is expected as the life time of dyon is exactly zero in the limit of infinite monopole mass.
Integrating over a $3$-ball around the monopole, one may explicitly check that the electric charge is conserved
\begin{equation}
  \int_{B^3} \dd^3 x\, g J_\text{el}^0 (t,r) + \int_0^t \dd \tau \int_{S^2} \dd \bm{S} \cdot \bm{J}_\text{el} (\tau,r) = 0.
\end{equation}
Because the first term tends to zero as the size of $B^3$ tends to zero, the net current on the infinitesimal surface around the monopole is zero, which means that the electric current never flows into the monopole, rather its charge is just surrounding the monopole. 
%This also implies that the electric current never flows into the monopole, rather its charge is just surrounding the monopole.
On the contrary, the chiral charge is generated from the monopole.
We can see this by again integrating over a $3$-ball around the monopole
\begin{equation}
  \varDelta Q_5 = 
  \int_{B^3} \dd^3 x\, J^0_5 (t,r) + \int_0^t \dd \tau \int_{S^2} \dd \bm{S} \cdot \bm{J}_5 (\tau,r)
  \simeq \frac{v_\theta t}{2 \pi} = \frac{\varDelta \theta (t)}{2 \pi}.
  \label{eq:q5_via_CReff}
\end{equation}
The unit chiral charge is generated every $\varDelta \theta (t) = 2 \pi n$.
As the topological density $q(x) \equiv g^2 \bm E\cdot \bm B/(4\pi^2)$ approaches to zero as $g\rightarrow 0$,
Eq.~\eqref{eq:q5_via_CReff} seems to contradict the anomaly $\partial_\mu J^\mu_5 = 2 q(x)$.
However, the anomaly equation is actually satisfied, because the topological density goes to zero only when $r>0$, and the correct limit as $g\to0$ including $r=0$ is
\begin{align}
  %\frac{g^2}{4\pi^2}\bm E\cdot \bm B 
  q(x) = - \frac{g^2(\phi-\theta)}{64\pi^3r^4} \to \frac{v_\theta}{4\pi}\delta^{(3)}(\bm x).
  \label{eq:topo_dens}
\end{align}
This limit is shown in Appendix~\ref{sec:A}.

We can interpret the origin of the chiral charge intuitively.
The only outgoing fermion that cancels the incoming $\bm{J}_\text{CS}$ is the right-handed particle $\psi_{R,s}$, and hence the monopole inevitably generates the chiral asymmetry to avoid the formation of dyon.
When the change of the $\theta$ term stops, the production of fermions at the core of the monopole terminates, but the generated chiral charges remain and propagate at the speed of light.
One might still wonder how come it is possible to cancel out the electric charge flow from $J_\text{CS}$ since the electric charge carried by particle excitations is quantized while the charge carried by $J_\text{CS}$ is continuous.
We will come back this subtlety in the last part of this section.

\paragraph{Quantum solution.}
Now we are in a position to quantize the boson theory.
As we have seen in the classical solution, the essential dynamics can be understood in the weak-coupling ($g \to 0$) limit.
There the chiral charge is generated at the location of the monopole while the emitted particle propagates freely.
For simplicity, we quantize the boson theory in this weak-coupling limit, whose action reads
\begin{equation}
  S_b = \int \dd t \int_0^{2L} \dd r\, \frac{1}{8 \pi} \qty[ \qty(\partial_t \phi)^2 - \qty(\partial_r \phi)^2 ],
\end{equation}
with the boundary condition being
\begin{equation}
  \phi = \theta \quad \text{at}~~~ r = 0, 2L.
  \label{boundary_cond}
\end{equation}
The boundary condition at $r = 2L$ reflects the Neumann boundary condition at $r = L$ in the classical solution.
Again, the validity of the solution is restricted to $r,t \ll L$.

The general solution of the equation of motion under the boundary condition is
\begin{equation}
  \label{eq:qsol_monopole_M}
  \hat\phi(t,r)=\sum_{n=1}^\infty 2i(\hat B_ne^{-i\bar k_n t} -\hat B_n^\dag e^{i\bar k_nt})\sin\bar k_n r + v_\theta t
\end{equation}
with $\bar k_n = n \pi / (2L)$.
The canonical commutation relation%
\footnote{
To obtain the commutation relation of $\hat B_n$, we use
\begin{align*}
  \frac1L\sum_{n=1}^\infty \sin\bar k_nr \sin \bar k_nr' = -\delta(r+r') + \delta(r-r'),
\end{align*}
where the delta functions are $4L$ periodic.
Here we can neglect the first term because $0<r+r'<4L$ and thus $\delta(r+r')=0$.
}%
$[\phi(t,r),\partial_t\phi(t,r')] = 4\pi i\delta(r-r')$ implies that the operators $\hat B_n$ has to satisfy $[\hat B_n,\hat B_m^\dag]=\delta_{nm}/n$.
Using these operators, the Hamiltonian is expressed as
\begin{equation}
  \hat H = \frac1{8\pi}\int_0^{2L}\dd x \, \qty[ \qty(\partial_t\hat\phi)^2 + \qty(\partial_r\hat\phi)^2]
  =\frac\pi{2L}\sum_{n=1}^\infty n^2 \big[ \hat B_n^\dag - X^*_n(t) \big] \big[\hat B_n - X_n(t) \big],
\end{equation}
where we define
\begin{align}
  X_n(t) \equiv \begin{dcases}
  0 & \text{for }n\in\text{even},\\
  -\frac{2Lv_\theta}{\pi^2n^2}e^{i\bar k_nt} & \text{for }n\in\text{odd}.
  \end{dcases}
\end{align}
We see that the Hamiltonian depends on time explicitly, and thus the eigenstates also depends on time.
Note that we are now in the Heisenberg picture, and thus states do not evolve with time.
The time-dependence of the eigenstate does not mean the time evolution, but only that the eigenstates at each time are different from each other.
The vacuum state is characterized by the equation
\begin{equation}
  \big[ \hat B_n - X_n(t) \big] \ket{\Omega(t)} = 0\quad\text{for all $n$},
\end{equation}
and the other eigenstates are obtained by acting the operators $\hat B_n^\dag - X^*_n(t)$ to it.
The vacua at different times are related by a unitary transformation,
%\ckm{I flipped the definition of $U$ so that it coincides with the conventional time evolution operator (*time evolution of eigenvector is backward in Heisenberg picture).}
\begin{equation}
  \hat U(t) \ket{\Omega(t)} = \ket{\Omega(0)},\quad 
  \hat U(t)=\exp\qty{ \sum_{n\in\text{odd}, n>0}\left[ - \alpha_n(t)\hat B_n^\dag + \alpha_n^*(t)\hat B_n\right] },
  \quad \alpha_n (t) \equiv \frac{2Lv_\theta}{\pi^2 n}\left(1-e^{i\bar k_n t}\right).
\end{equation}
This equation means that the vacuum at a given time is a coherent state with respect to the vacuum at a different time.

The classical solution (\ref{eq:monopole_sol_aprx}) is reproduced by taking the expectation value with respect to the vacuum $\ket{\Omega(0)}$ at $t=0$.
One may confirm this immediately by rewriting Eq.~\eqref{eq:qsol_monopole_M} in the following form:
\begin{equation}
  \hat\phi (t, r) = \sum_{n \in \mathbb{Z}_+} 2 i \qty( \hat b_n e^{- i \bar k_n t} - \hat b_n^\dag e^{i \bar k_n t}) \sin {\bar k_n r}
  + \phi_R (t-r) + \phi_L (t + r),
  \label{eq:qsol_monopole}
\end{equation}
with $\hat b_n \equiv \hat B_n - X_n (0)$, \textit{i.e.}, $\hat b_n \ket{\Omega (0)} = 0$.
%We add a hat to operators occasionally when we would like to distinguish them from $c$-numbers.
$\phi_R$ and $\phi_L$ are $c$-number functions, which are given by
\begin{equation}
  \label{eq:classical_with_tf}
  \phi_R (t-r) = 
  \begin{cases}
    0 & \text{for}~~ t-r < 0 \\
    v_\theta \qty( t - r ) & \text{for}~~ 0 \leqslant t - r %< t_f \\
    %v_\theta t_f & \text{for}~~  t_f \leqslant t - r
\end{cases}, \qquad
\phi_L (t + r) = \phi_R (t - 2 L + r),
\end{equation}
for $t \ll L$.
Hence, the classical solution \eqref{eq:monopole_sol_aprx} is reproduced as $\langle \hat \phi \rangle$ for $0 \leqslant r \leqslant L$.
This is reasonable because the initial condition of our solution is $\phi=\partial_t\phi=0$, which minimizes the energy at $t=0$.

In the basis $\hat B_n-X_n(t_0)$ diagonalizing the Hamiltonian at $t=t_0$, the currents are written as
\begin{align}
  \hat J^0_{\bm 2}(t,r) &= \hat J^r_{\bm 2,5}(t,r) = -\partial_r \phi(t,r)/(2\pi)\notag\\
  &= -\frac1{2L}\sum_{n=1}^\infty \qty{ in \qty[\hat B_n-X_n(t_0)]e^{-i\bar k_n t} +\text{H.c.} } \cos\bar k_n r - \frac1{2\pi}\partial_r\qty[\phi_R(t-t_0-r) + \phi_L(t-t_0+r)],\notag\\
  \hat J^0_{\bm 2,5}(t,r) &= \hat J^r_{\bm 2}(t,r) = \partial_t \phi(t,r)/(2\pi)\notag\\
  &= \frac1{2L}\sum_{n=1}^\infty \qty{ n \qty[ \hat B_n-X_n(t_0)]e^{-i\bar k_n t}+\text{H.c.}}\sin\bar k_n r + \frac1{2\pi}\partial_t\qty[\phi_R(t-t_0-r) + \phi_L(t-t_0+r)].
  \label{current_boson}
\end{align}
We see that the vacuum expectation values of the currents at $t=t_0$ are zero because $\phi_R'(-r) = \phi_L'(r) = 0$.

\paragraph{Correspondence to the fermion theory}

To clarify the interpretation of the states in the bosonized theory, we solve the fermion theory and look at the operator correspondence.
The action is
\begin{align}
  S_f = \int \dd t\int_0^{2L}\dd r \qty[ i\psi_L^\dag \qty(\partial_t-\partial_r)\psi_L + i\psi_R^\dag \qty(\partial_t+\partial_r)\psi_R].
\end{align}
The boundary condition corresponding to Eq.~(\ref{boundary_cond}) is
\begin{align}
  e^{i\theta}\hat\psi_R|_{r=0}=\hat\psi_L|_{r=0} ,\quad
  e^{i\theta}\hat\psi_R|_{r=L} = -\hat\psi_L|_{r=L}.
\end{align}
We can explicitly show that this theory is equivalent to the bosonized theory by comparing the spectrum.
As we have seen in the bosonized theory, the vacuum depends on time.
The general solution as a series around the vacuum at $t=t_0$ is given as
\begin{align}
  \hat\psi_{R/L}(t,r) = \frac1{\sqrt{4L}}\sum_{n\in\mathbb Z}\hat a_n(t_0) e^{-ik_n(t\mp r)\mp i\phi_{R/L}(t-t_0+r) \mp iv_\theta t_0/2},
  \label{expansion}
\end{align}
where the momentum is $k_n=\pi(n+1/2)/(2L)$.
The canonical anti-commutation relation implies that the creation/annihilation operators fulfill $\{\hat a_n(t_0),\hat a_m^\dag(t_0)\}=\delta_{nm}$.
The Hamiltonian is diagonalized in this basis at $t=t_0$, and the vacuum $\ket{\Omega(t_0)}$ at $t=t_0$ is characterized by
\begin{align}
  \hat a_n(t_0)\ket{\Omega(t_0)} = \hat a_{-n-1}(t_0)\ket{\Omega(t_0)} = 0,\quad \text{for }n\in\mathbb Z_{\geq 0}.
\end{align}
The creation/annihilation operators at different times are related by a unitary transformation, which is shown in Appendix \ref{sec:appendix_monopole}.

To see the correspondence between the creation/annihilation operators of fermions and bosons, let us compare the expressions of the currents.
In the fermion theory, we have to fix the c-number part of the currents, which depends on the choice of the regularization.
The proper choice has to respect the "charge conjugation" symmetry
\begin{align}
  \hat\psi_L(t,r) \to \hat C(t)\hat\psi_L(t,r)\hat C^\dag(t) = e^{i\theta(t)} \hat\psi_L^\dag(t,r),\quad \hat\psi_R(t,r)\to \hat C(t)\psi_R(t,r)\hat C^\dag(t) = e^{-i\theta(t)}\hat\psi_R^\dag,
  \label{charge_conjugate_def}
\end{align}
which means that, under this transformation, the currents have to flip their signs.
At $t=t_0$, the transformation acts on the creation/annihilation operators $a_n(t_0)$ as
\begin{align}
  \hat C(t_0)\hat a_n(t_0)\hat C^\dag(t_0) = \hat a^\dag_{-n-1}(t_0).
  \label{c_at0}
\end{align}
Note that $\hat C(t)$ acts on $\hat a_n(t_0)$ differently when $t\neq t_0$.
In the basis of Eq.~(\ref{expansion}), the currents are written as
\begin{align}
  {:}\hat\psi_{R/L}^\dag(t,r)\hat\psi_{R/L}(t,r){:}
  &=\frac1{4L}\sum_{n=1}^{\infty}\qty[\sum_{m\in\mathbb Z}\hat a_n^\dag(t_0)\hat a_{n+m}(t_0) e^{-i\bar k_n(t\mp r)} + \text{H.c.}]\notag\\
  &\quad+ \lim_{\varepsilon\to0}\frac1{4L}\sum_{n\in\mathbb Z}e^{-\varepsilon|k_n|}\qty[\hat a^\dag_n(t_0)\hat a_n(t_0) - 1/2 ] + c_{R/L}(t,r;t_0),
\end{align}
where $c_{R/L}(t,r;t_0)$ is the c-number part.
Here we choose the regulator as $\exp(-\varepsilon|k_n|)$, and when we change this, the c-number part $c_{R/L}(t,r;t_0)$ changes.
The c-number part $c_{R/L}(t,r;t_0)$ should be chosen so that
\begin{align}
  \hat C(t) {:}\hat\psi_{R/L}^\dag(t,r)\hat\psi_{R/L}(t,r){:} \hat C^\dag(t) = 
  -{:}\hat\psi_{R/L}^\dag(t,r)\hat\psi_{R/L}(t,r){:}
  \label{c_ql}
\end{align}
Substituting Eq.~(\ref{c_at0}) into Eq.~(\ref{c_ql}) at $t=t_0$, we obtain
\begin{align}
  c_{R/L}(t=t_0,r;t_0) = 0.
  \label{c_t0}
\end{align}
Note that at this stage we have not yet determined $c_{R/L}(t,r;t_0)$ for $t\neq t_0$, which will be determined in Appendix \ref{sec:appendix_monopole}.
Thus the expression of the currents at $t=t_0$ is obtained as
\begin{align}
  \hat J^0_{\bm 2}(t_0,r) &= \hat J^r_{\bm 2,5}(t_0,r) = {:}\hat\psi_L^\dag\hat\psi_L{:}+{:}\hat\psi_R^\dag\hat\psi_R{:}\notag\\
  &=\frac1{2L}\sum_{n=1}^{\infty}\qty[\sum_{m\in\mathbb Z}\hat a_m^\dag(t_0)\hat a_{m+n}(t_0) e^{-i\bar k_n t_0} +\text{H.c.}]\cos\bar k_nr
  +\lim_{\varepsilon\to0}\frac1{2L}\sum_{n\in\mathbb Z}e^{-\varepsilon|k_n|}\qty[\hat a^\dag_n(t_0)\hat a_n(t_0) - 1/2],\notag\\
  \hat J^0_{\bm2,5}(t_0,r) &= \hat J^r_{\bm 2}(t_0,r) = -{:}\hat\psi_L^\dag\hat\psi_L{:} + {:}\hat\psi_R^\dag\hat\psi_R{:}\notag\\
  &=\frac1{2L}\sum_{n=1}^{\infty}\qty[i\sum_{m\in\mathbb Z}\hat a_m^\dag(t_0)\hat a_{m+n}(t_0) e^{-i\bar k_n t_0} +\text{H.c.}]\sin\bar k_nr.
  \label{current_fermion}
\end{align}
By comparing with the expression (\ref{current_boson}) using the boson field, we see the correspondence
\begin{align}
  -in \qty[\hat B_n - X_n(t_0)] = \sum_{m\in\mathbb Z}\hat a_{m}^\dag(t_0)\hat a_{m+n}(t_0),
  \label{correspondence}
\end{align}
which satisfies the required property of $[\hat B_n, \hat B_m^\dag] = \delta_{nm}/n$.
Note that the bosonized theory only expresses the $Q_V=0$ sector of the fermion theory\footnote{
The $Q_V = n$ sector corresponds to the bosonized theory with the boundary condition $\phi|_{x=L}=\theta + 2\pi n,\ \phi|_{x=0}=\theta$.}, and there are no terms corresponding to the second term of $\hat J^0_{\bm 2}$ in Eq.~(\ref{current_fermion}).
In the Appendix, we show that the spectrum actually matches between the fermion and boson theories.

As seen in the bosonized theory, the vacuum $\ket{\Omega(0)}$ at $t=0$ is regarded as a coherent state with respect to the vacuum $\ket{\Omega(t)}$ at the different time $t\neq 0$.
From the correspondence (\ref{correspondence}), we see that the one-boson state is an entangled two-fermion state with zero charge.
Therefore the classical solution of Eq.~\eqref{eq:quantities} should be interpreted as an expectation value with respect to a multi-body state of particle and anti-particle.
This also implies that the non-integer charge emission from a monopole for $\theta \neq 2 \pi n$ is originated from a skewed distribution of entangled particle and anti-particle pairs in the wave function.
We will explicitly confirm this expectation in the later discussion.

\paragraph{Interpretation.}
Now we have the time-evolution of the observables and the state corresponding to our situation, which in principle enable us to answer any question.
As we have confirmed explicitly, all the operators fulfill the required conservation laws and therefore the expectation values of any operators does.
The confusions arise only when we come to discuss the observation of individual fermions.
In particular, the main question is whether fermions are actually observed or not.

\begin{figure}[t]
\centering
\includegraphics[width=0.7\hsize]{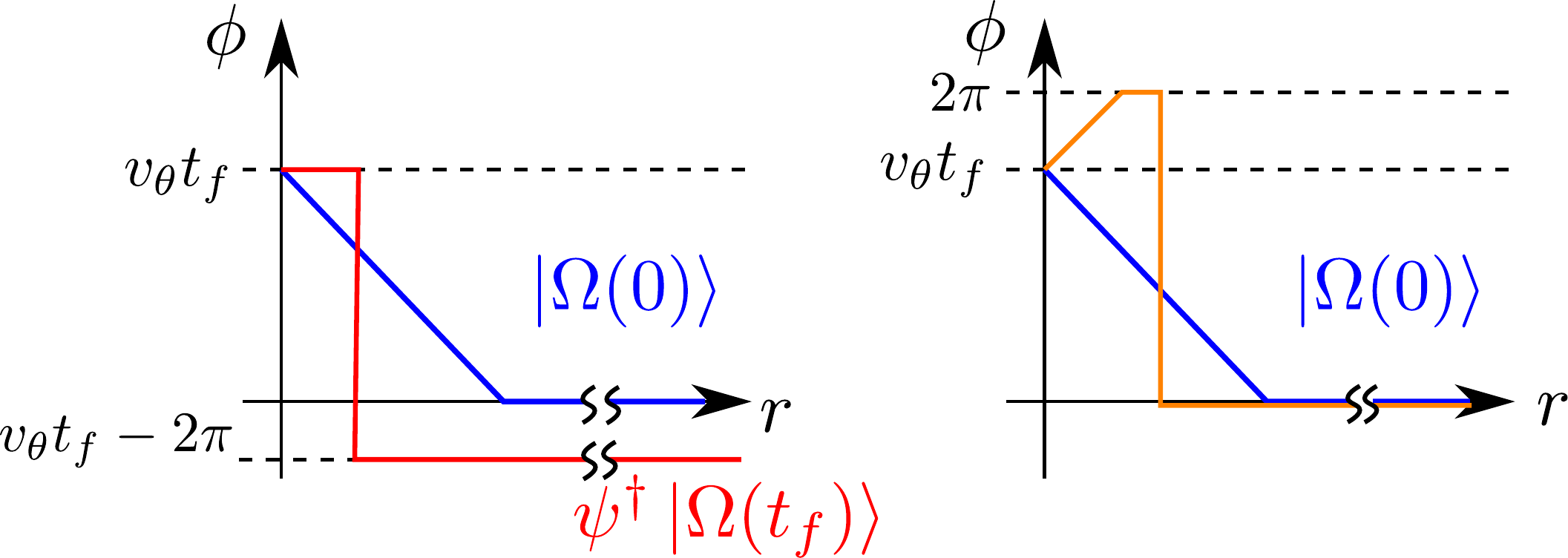}
\caption{The profile of $\phi$. The blue line corresponds to the vacuum $\ket{\Omega(0)}$ at $t=0$. The red line corresponds to the one-particle state $\psi^\dag\ket{\Omega(t_f)}$ at $t=t_f$. In the bosonized theory, the local creation of a fermion corresponds to the insertion of a sharp kink.
Due to the large misalignment between the two profiles, the states are orthogonal to each other in $L\to\infty$ limit. The orange line represents an example of the profiles that can be regarded as a one-particle state locally and have a finite overlap with $\ket{\Omega(0)}$.}
\label{fig_phi_prof}
\end{figure}

To clarify this, we calculate the probability to observe a two-particle state, where a fermion propagates from $r=0$ and an anti-fermion propagates from $r=2L$.
Usually, a multiple-particle state is defined as a Fock state.
As we have seen, the Hamiltonian depends on time, and thus the Fock space also depends on time.
To confirm that we can observe particles, it seems to be sufficient to use the Fock space at the time $t=t_f$ when we perform an observation.
However, the overlap with the Fock space approaches zero as $L\to\infty$ except for $t_f=2\pi m/v_\theta$ for $m\in\mathbb Z$, which is calculated in Appendix \ref{sec:appendix_probability}.
Does this mean the particle are observed only when $t_f=2\pi m/v_\theta$ exactly?
The answer is no.
When we use a Fock space as a multi-particle state, we implicitly assume that the space outside the detector is the vacuum.
However, our state $\ket{\Omega(0)}$ is largely different from the vacuum and thus we should not use this definition of the multi-particle state.
Note that when we observe the particle locally, no information is available for locations where there is no detector, and thus a local observation does not confirm that the space is almost entirely vacuum.
There are states that can be regarded as a one-particle state locally and have a finite overlap with $\ket{\Omega(0)}$.
See Fig.~\ref{fig_phi_prof}.
By comparing the profile of the expectation value of $\phi(x)$, we can understand why our state $\ket{\Omega(0)}$ is orthogonal to the Fock space.
At $t=t_f$, the vacuum state should satisfy $\phi=v_\theta t_f \bmod 2\pi$, while $\phi=0$ in the most of the space for our state $\ket{\Omega(0)}$.
Thus the two profiles of $\phi$ are largely misaligned.
We can easily write a profile of $\phi$ that is locally similar to a Fock state, but it is similar to $\ket{\Omega(0)}$ in the other region of the space.
Such a state is expected to have nonzero overlap with $\ket{\Omega(0)}$.
In Appedix \ref{sec:appendix_probability}, we give an explicit example of such state.

%%%%%%%%%%%%%%%%%%%%%%%%%%%%%%%%%%%%%%%%%%%%%%%%%%
\section{Magnetic flux}
\label{sec:flux}

In this section, we study the chiral asymmetry production in the background of a magnetic flux (loop).
In the presence of a monopole, the chiral asymmetry is %directly
generated from %the boundary of a half line, \textit{i.e.}, at the location of 
the monopole as we have shown in Sec.~\ref{sec:monopole}.
In the case of a magnetic flux, however, we expect that the asymmetry generation takes place in a somewhat different manner since there are no singularities.

The main purpose of this section is to see how the time-evolution of $\theta$ leads to the chiral asymmetry generation.
We first derive the effective two-dimensional theory by focusing on the lowest Landau level and then bosonize it.
We demonstrate the production of the chiral charge \textit{quantum mechanically} by explicitly solving the Schr\"{o}dinger equation.
This example is not only theoretically interesting by itself, but provides interesting baryogenesis mechanism in a certain class of axion inflation models.

\subsection{Bosonization of the lowest Landau level}
\label{subsec:bosonized_flux}
Throughout this section, we consider the background of a homogeneous magnetic flux along the $z$-axis, whose vector potential can be expressed as
\begin{equation}
  \overline{A} = - B x \dd y.
  \label{eq:flux}
\end{equation}
For later convenience, we put this configuration on a $3$-torus: $T^3 \equiv \{ (x,y,z)| x \sim x+L, y \sim y+L, z \sim z+L \}$.
As long as the amplitude of the magnetic field is much larger than any physical scale of our interest, this simplified assumption can be justified a posteriori even in a more realistic situation discussed in Sec.~\ref{sec:concl}.
To identify the gauge field at $x$ with $x + L$, the difference should be compensated by a gauge redundancy of $\overline{A} \sim \overline{A} - \dd \alpha/g$ with $\alpha = g B L y$.
The gauge parameter $\alpha$ is uniquely determined on $T^3$ modulo $2 \pi$, \textit{i.e.}, $2 \pi N = \alpha|_{y+L} - \alpha|_y = g B L^2$, which implies that the number of magnetic flux penetrating the $(x,y)$-plane should be an integer, \textit{i.e.}, $- (g / 2 \pi)\int_{\bm{x},\bm{y}} \dd \overline A = g B L ^2 / 2 \pi = N$.

In the following, we only consider the dynamics along the magnetic field, \textit{i.e.}, the lowest Landau level, by assuming that any physical scale of our interest is much smaller than the amplitude of the magnetic field.
Under this approximation, the gauge field configuration can be expressed as
\begin{equation}
  A = A_t (t,z) \dd t + A_z (t,z) \dd z + \overline{A},
\end{equation}
and the fermion field is
\begin{equation}
  \psi = \frac{1}{\sqrt{L}} \sum_{f\in\mathbb Z} e^{i k_{f} y} h_0 \qty( x - x_f ) \chi_+^z \otimes 
  \begin{pmatrix}
    \psi_{L,f} \qty( t,z ) \\
    \psi_{R,f} \qty( t,z )
  \end{pmatrix}, \quad
  \chi_+^z \equiv 
  \begin{pmatrix}
    1 \\ 0
  \end{pmatrix},
  \quad
  h_0 \qty(x) \equiv \qty( \frac{g B}{\pi} )^{\frac{1}{4}} e^{ - \frac{g B}{2}x^2 }.
\end{equation}
%Here we define $x_f\equiv Lf/N$ and $k_f  = \pi (2 f + 1) / L $ so that the fermion field fulfills the anti-periodic boundary condition.
Here we define $x_f\equiv Lf/N$ and $k_f=\pi(2f+1)/L$. The anti-periodic boundary condition implies%
\footnote{
By summing over $f=k\bmod N$, we obtain the coefficient of $\psi_{R/L,k}$ for a specific $k$, and we see that the shift symmetry for $x,y$ is broken.
This is not problematic because the background field (\ref{eq:flux}) explicitly breaks the shift symmetry, even though the magnetic field $B$ itself does not break.
The breaking in the $x$ direction is seen in the Wilson loop along the $y$ direction.
The breaking in the $y$ direction is seen in the gauge transformation function $\lambda(y)= -BLy$, which relates the values of the gauge field at $x=0$ and $x=L$ as $A(x=L)-A(x=0)=\dd \lambda(y)$.
This is one of the differences from the monopole case, where the rotational symmetry is not broken.
}
$\psi_{R/L,f} = -\psi_{R/L,f+N}$.
In this particular gauge \eqref{eq:flux}, the anti-periodic boundary condition for $x \sim x + L$ is guaranteed up to the gauge redundancy of $( \psi, A ) \sim (e^{i g B L y} \psi, A  - B L \dd y)$. 

Inserting these configurations to the original action, we obtain the effective action for the lowest Landau level
\begin{align}
  S &= \int \dd t \int_0^L \dd z\, \sum_{f = 1}^N \qty[ 
    \frac{1}{g B / (2 \pi)} \frac{1}{2} F_{tz}^2 
    + i \psi_{L,f}^\dag \qty( D_t - D_z ) \psi_{L,f} + i \psi_{R,f}^\dag \qty( D_t + D_z ) \psi_{R,f} 
    - \frac{g \theta}{2 \pi} F_{tz}
    ] \\
    &= \int \dd t \int_0^L \dd z\, \sum_{f = 1}^N \qty[
    \frac{1}{g B / (2 \pi)} \frac{1}{2} F_{tz}^2 
    + \overline\psi_{\bm{2},f} i \gamma_{\bm{2}}^\mu D_\mu \psi_{\bm{2},f}
    - \frac{g \theta}{2 \pi} F_{tz}
    ].
\end{align}
In the second line, we introduce two-dimensional Dirac fields as $\psi_{\bm{2},f} \equiv ( \psi_{L,f}, \psi_{R,f} )^T$.
One may readily see that the effective action describes a two-dimensional theory on a segment of $[0,L]$, involving left- and right-handed fermions on each magnetic flux.
This theory has a classical symmetry of $U (N)_R \times U(N)_L \simeq U(1)_V \times U(1)_A \times SU(N)_V \times SU(N)_A$, whose
corresponding currents are given by
\begin{equation}
    J_{\bm{2}}^\mu \equiv \sum_f \overline{\psi}_{\bm{2},f} \gamma_{\bm{2}}^\mu \psi_{\bm{2},f}, \quad
    J_{\bm{2},5}^\mu \equiv \sum_f \overline{\psi}_{\bm{2},f} \gamma_{\bm{2}}^\mu \gamma_{\bm{2}}^5 \psi_{\bm{2},f}, \quad
    J_{\bm{2}}^{a \mu} \equiv \sum_{ij} \overline{\psi}_{\bm{2},i} \gamma_{\bm{2}}^\mu T^a_{ij} \psi_{\bm{2},j}, \quad
    J_{\bm{2},5}^{a \mu} \equiv \sum_{ij} \overline{\psi}_{\bm{2},i} \gamma_{\bm{2}}^\mu \gamma_{\bm{2}}^5 T^a_{ij} \psi_{\bm{2},j}.
\end{equation}
These two-dimensional currents are related to their four-dimensional counterparts via $J^\mu = L^{-2} J^\mu_{\bm{2}}$, $J^\mu_5 = L^{-2} J^\mu_{\bm{2},5}$ and the same relations for non-Abelian currents.
The non-Abelian part is an emergent symmetry at the lowest Landau level.

Two-dimensional fermions with such a non-Abelian symmetry can also be bosonized~\cite{Witten:1983ar}.
The Abelian currents are described by the phase $\phi$ in the same way as the previous section, namely
\begin{align}
  \label{eq:vector2d_flux}
  J_{\bm{2}}^t &= - \frac{N}{2\pi} \partial_z \phi, \quad   J_{\bm{2}}^z = \frac{N}{2\pi} \partial_t \phi, \\
  \label{eq:chiral2d_flux}
  J_{\bm{2},5}^t &= \frac{N}{2\pi} \partial_t \phi, \quad ~ J_{\bm{2},5}^z = - \frac{N}{2\pi} \partial_z \phi,
\end{align}
which fulfills the required property of $J_{\bm{2}}^\mu = - \epsilon^{\mu\nu} J_{\bm{2},\nu}$.
On the other hand, the non-Abelian currents are expressed by an $SU(N)$ matrix boson field, 
$\sigma: M^2 \to SU(N)$, as
\begin{align}
  J_{\bm{2}}^{at} = \frac{i}{4\pi} \qty{ \tr \qty[ [\sigma^\dag, \partial_t \sigma]_- T^a ] + \tr \qty[ [\sigma^\dag, \partial_z \sigma]_+ T^a ]   }, \quad 
  J_{\bm{2}}^{az} = -\frac{i}{4\pi} \qty{ \tr \qty[ [\sigma^\dag, \partial_t \sigma]_+ T^a ] + \tr \qty[ [\sigma^\dag, \partial_z \sigma]_- T^a ]   },
\end{align}
and its axial counterpart is obtained from $J_{\bm{2},5}^{a\mu} = - \epsilon^{\mu\nu} J_{\bm{2},\nu}^a$.
The bosonized action is given as a functional of $\phi$ and $\sigma$:
\begin{equation}
  S_b = N \int \dd t \int_0^L \dd z\, \qty{ \frac{1}{gB/ (2 \pi)} \frac{1}{2} F_{tz}^2 + \frac{g}{2\pi} \qty( \phi - \theta ) F_{tr}
  + \frac{1}{8 \pi} \qty[ \qty(\partial_t \phi)^2 - \qty(\partial_z \phi)^2 ]}
  + S_\text{WZW}(\sigma),
\end{equation}
where $S_\text{WZW} (\sigma)$ represents the Wess-Zumino-Witten model of the level one.

Again, we can further simplify the action by solving the equation of motion for the gauge field
\begin{equation}
  \partial_z F_{tz} + \frac{g B}{2 \pi} \frac{g}{2 \pi} \partial_z \qty( \phi - \theta ) = 0, \qquad
  \partial_t F_{tz} + \frac{g B}{2 \pi} \frac{g}{2 \pi} \partial_t \qty( \phi - \theta ) = 0.
  \label{eq:gauss_flux}
\end{equation}
The general solution can be written as
\begin{equation}
  F_{tz} = - \frac{g B}{2 \pi} \frac{g}{2 \pi} \qty( \phi - \theta + c ),
  \label{eq:Ftz}
\end{equation}
where the integration constant $c$ is determined so that the electric field vanishes initially, $t = 0$.
Absorbing $c$ into the definition of $\theta$, we therefore enforce $\phi = \theta$ for $t \leqslant 0$.
Eventually, we obtain the following action
\begin{equation}
  S_b = N \int \dd t \int_0^L \dd z\, \qty{
    \frac{1}{8 \pi} \qty[ ( \partial_t \phi )^2 - ( \partial_z \phi )^2 ] -
    \frac{1}{2} \frac{g B}{2 \pi} \qty( g \frac{\phi - \theta}{2 \pi} )^2
  }
  + S_\text{WZW}(\sigma).
\end{equation}
Note that the $2\pi$ periodicity is satisfied as argued around Eq.~\eqref{eq:action_boson}.
For massless fermions, $\phi$ and $\sigma$ are decoupled and hence we omit $S_\text{WZW} (\sigma)$ hereafter.

\subsection{Production of chiral asymmetry along the flux}
\label{subsec:q5_flux}

\paragraph{Classical solution.}
Let us again start with a classical solution.
The equation of motion for $\phi$ reads
\begin{equation}
  \partial_t^2 \phi - \partial_z^2 \phi + \frac{g B}{2 \pi} \frac{g^2}{\pi} \qty( \phi - \theta ) = 0.
  \label{eq:eom_flux}
\end{equation}
Again the anomaly equation is automatically satisfied as one can see from Eqs.~\eqref{eq:vector2d_flux}, \eqref{eq:chiral2d_flux}, and \eqref{eq:Ftz}.

Suppose that $\theta$ starts to evolve at $t = 0$.
Our initial condition does not have any net electric nor chiral charge, \textit{i.e.}, $\phi = 0$, $\dot \phi = 0$ for $t < 0$.
The solution under this initial condition is readily obtained with a help of the retarded Green function
\begin{equation}
  \phi (t) = \omega_B \int^t_0 \dd t' \sin \qty[ \omega_B (t - t') ] \theta (t'), \quad
  \omega_B^2 \equiv \frac{g^3 B}{2\pi^2}.
\end{equation}
Inserting the time evolution of $\theta$ in Eq.~\eqref{eq:theta_t}, one may perform the integration, which reads
\begin{equation}
  \label{eq:flux_sol_cl}
  \phi (t) = v_\theta t - \frac{v_\theta}{\omega_B} \sin \omega_B t.
\end{equation}
From this, we obtain the time evolution of any current and the electric field
\begin{equation}
  \label{eq:flux_sol_cl_J}
  J^0 = 0, \quad
  J^z = \frac{g B}{2 \pi} \frac{v_\theta}{2 \pi} \qty( 1 - \cos \omega_B t ),
  \quad
  J_5^0 = \frac{g B}{2 \pi} \frac{v_\theta}{2 \pi} \qty( 1 - \cos \omega_B t ),
  \quad
  J_{5}^z = \bm{0}, 
  \quad
  E_z = - \frac{g B}{2 \pi} \frac{g v_\theta}{2 \pi \omega_B} \sin \omega_B t.
\end{equation}
The current from the Chern-Simons coupling is given by
\begin{equation}
  J_\text{CS}^z = - \frac{g B}{2 \pi} \frac{g v_\theta}{2 \pi},
\end{equation}
and hence the electric current is oscillating around zero
\begin{equation}
  \bm{J}_\text{el} = g \bm{J} + \bm{J}_\text{CS} 
  = 
  - \hat{\bm{z}} \frac{g B}{2 \pi} \frac{g v_\theta}{2 \pi} \cos \omega_B t.
\end{equation}
This implies that the fermion current is induced to cancel out the CS current.

The amount of chiral charge generated via this process is obtained by integrating $J_5^0$ over $T^3$
\begin{equation}
  \label{eq:Q5_flux_cl}
  \Delta Q_5 = \int_{T^3} J_5^0 
  = 2 N \times \frac{v_\theta/2}{2 \pi} L \times \qty( 1 - \cos \omega_B t ).
\end{equation}
Contrary to the case of a monopole background, the production of the net chiral charge occurs neither proportional to $\theta$ nor in a unit of $\varDelta \theta = 2 \pi n$.
Let us take a closer look at the origin of the chiral charge to understand the underlying physics intuitively.
As can be seen immediately from our starting point \eqref{eq:setup}, a time-dependent $\theta$ in turn implies a chemical potential imposed on the chiral charge as $(\dot \theta / 2) J^0_5$.
The Fermi momentum changes according to this chemical potential, \textit{i.e.} $\Delta p_F = \pm v_\theta / 2$ for the right- /left-handed fermions, because the chiral symmetry is broken by the quantum anomaly.
One may count the number of right-/left-handed fermions at each magnetic flux by integrating the density $\Delta p_F / (2 \pi)$, namely
$
  \oint \dd z\, \varDelta p_F/(2 \pi) = \pm L (v_\theta/2) / (2 \pi).
$
By summing over all the fluxes, we reproduce the result \eqref{eq:Q5_flux_cl} aside from the oscillating factor 
\begin{equation}
  N \times \qty( \frac{v_\theta / 2}{2 \pi}L - \frac{-v_\theta / 2}{2 \pi}L  ) = 2 N \times \frac{v_\theta/2}{2 \pi} L.
  \label{eq:intuition_flux}
\end{equation}
Once chiral fermions are generated, they form a current \textit{\`{a} la} the chiral magnetic effect.
Then the electric fields that decelerate fermions are induced, thereby reduce the Fermi momentum.
Such a backreaction with a negative feedback explains the overall oscillating factor in Eq.~\eqref{eq:Q5_flux_cl}.
On the other hand, in the case of a monopole background under the weak coupling limit, the chiral charge is only broken at the boundary by the time-dependent $\theta$.
In addition, under the restriction of $t \ll L$, chiral fermions generated at $r=0$ and $r = 2L$ never meet each other,
which implies the absence of the backreaction.
Hence, the chiral asymmetry is simply determined by $\theta / (2 \pi)$ without any oscillatory behavior.

Still, it is not clear what is responsible for a non-integer chiral charge when the Fermi momentum does not change by a unit, \textit{i.e.}, $\Delta p_F \neq 2 \pi n /L$.
Also a non-integer chiral charge is puzzling, in particular, when someone observes the number of fermions.
These subtleties will be addressed later by comparing the quantum solution derived in the next subsection to the original fermion theory.

\paragraph{Quantum solution.}
As we have seen in the previous section, the dynamics can be understood by the zero mode of $\phi$, whose action is given by
\begin{equation}
  S_b = N L \int \dd t\, \qty[ \frac{1}{8 \pi} \qty( \partial_t \phi_0 )^2 - \frac{1}{2} \frac{gB}{2 \pi} \qty(g \frac{\phi_0 - \theta}{2 \pi} )^2 ].
\end{equation}
Here the zero mode is defined by $\phi_0 \equiv \oint \dd z\, \phi / L$.

Let us start with the expression similar to Sec.~\ref{sec:monopole}
\begin{equation}
  \label{eq:qsol_flux_M}
    \hat \phi_0 (t) = \sqrt{\frac{2 \pi}{N L \omega_B}}  \qty( \hat B_0  e^{- i \omega_B t} + \text{H.c.} ) + v_\theta t,
\end{equation}
which solves the equation of motion for $t > 0$.
The creation/annihilation operators fulfill $[\hat B_0, \hat B_0^\dag] = 1$ that comes from the canonical commutation relation of $[\hat \phi_0, N\dot{\hat \phi}_0 / (4 \pi)] = i / L$.
The Hamiltonian in this basis reads
\begin{equation}
  H 
  = NL \qty[ \frac{\dot{\hat\phi}_0^2}{8 \pi} + \frac{1}{2} \frac{gB}{2 \pi} \qty( g \frac{\hat \phi_0 - \theta}{2 \pi} )^2 ]
  = \omega_B \big[ \hat B_0^\dag - Y^\ast (t) \big] \qty[ \hat B_0 - Y (t)],
\end{equation}
where 
\begin{equation}
  Y (t) = \frac{v_\theta}{2 i} \sqrt{\frac{NL}{2 \pi \omega_B}} e^{i \omega_B t}.
\end{equation}
The vacuum state at a time $t$ is defined by
\begin{equation}
  \qty[ \hat B_0 - Y (t) ] \ket{ \Omega (t) } = 0.
\end{equation}
Again, the vacua at different times are related through a unitary transformation as follows
\begin{equation}
  \hat U (t) \ket{ \Omega (t) } = \ket{\Omega (0)}, \quad 
  \hat U(t) = \exp \qty[ - \beta^\dag (t) \hat B_0 + \beta (t) \hat B_0^\dag ], \quad
  \beta (t) \equiv \frac{v_\theta}{2 i} \sqrt{ \frac{NL}{2 \pi \omega_B} } \qty( 1 - e^{i \omega_B t} ),
\end{equation} 
which implies that the vacuum at a certain time $t$ is a coherent state for another vacuum at a different time.

The classical solution \eqref{eq:flux_sol_cl} is reproduced if we take the expectation value with respect to $\ket{\Omega (0)}$,
which implies that the state corresponding to our initial condition minimizes the energy at $t = 0$ as expected.
One may see this explicitly by rewriting Eq.~\eqref{eq:qsol_flux_M} as follows
\begin{equation}
  \hat \phi_0 (t) = \sqrt{\frac{2 \pi}{N L \omega_B}}  \qty( \hat b_0 e^{- i \omega_B t} + \text{H.c.} ) + \phi (t),
  \label{eq:qsol_flux_b}
\end{equation}
with $\hat b_0 = \hat B_0 - Y (0)$, \textit{i.e.}, $\hat b_0 \ket{\Omega (0)} = 0$.
$\phi$ stands for a $c$-number function, which is given as
\begin{equation}
    \phi (t) = \begin{cases}
      0 & \text{for}~~~ t \leqslant 0, \\
      v_\theta t - \frac{v_\theta}{\omega_B} \sin \omega_B t & \text{for}~~~ 0 < t. %\leqslant t_f \\
      %v_\theta t_f - \frac{v_\theta}{\omega_B} \qty[ \sin \omega_B t - \sin \omega_B \qty( t - t_f ) ] & \text{for}~~~ t_f < t.
  \end{cases}
\end{equation}
It is obvious that the classical solution \eqref{eq:flux_sol_cl} is reproduced as $\langle \hat \phi_0 \rangle$.

\paragraph{Correspondence to the fermion theory.}
In the case of a monopole background in the weak coupling limit discussed in Sec.~\ref{sec:monopole}, 
the corresponding boson is free in the bulk and the breaking of the chiral symmetry is coming from the boundary.
Hence the relation between the boson and the original fermion is rather simple as given in Eq.~\eqref{correspondence}.
In contrast, in the current example, \textit{i.e.}, the background of a magnetic flux loop, 
the zero mode of the gauge field is present, which breaks the chiral symmetry everywhere.
This makes the relation between the boson theory and the original fermion theory somewhat complicated.

For clarity, let us take a closer look at this relation, starting from the original fermion theory:
\begin{align}
        S =& \int \dd t\, \Bigg\{ N L  \qty( \frac{1}{g B / 2 \pi} \frac{\dot A_z^2}{2} - \frac{g \theta}{2 \pi} \dot A_z )
        - g A_z \sum_{f = 1}^N  \oint \dd z\, \qty( \psi^\dag_{R,f} \psi_{R,f} - \psi^\dag_{L,f} \psi_{L,f} ) \\
        &+ \sum_{f = 1}^N \oint \dd z\, \qty[ i \psi_{R,f}^\dag \qty( \partial_t + \partial_z ) \psi_{R,f} + i \psi_{L,f}^\dag \qty( \partial_t - \partial_z ) \psi_{L,f}  ] \Bigg\},
\end{align}
with the boundary condition being $\psi_{R/L,j} |_{z = 0} = - \psi_{R/L, j} |_{z = L}$.
The chiral charge is broken everywhere as the chiral fermions are interacting with a zero mode of the gauge field $A_z (t)$.
Nevertheless, one may exactly solve the Dirac equation,
\begin{equation}
 i (\partial_t \pm \partial_z)\psi_{R/L, f} \mp g A_z \psi_{R/L,f}=0,
\end{equation}
 under an arbitrary background of $A_z (t)$ as
\begin{equation}
    \hat\psi_{R/L} = \frac{1}{\sqrt{L}} \sum_{j \in \mathbb{Z}} \hat a_j^{R/L} e^{ - i k_j \qty(t \mp z) \mp i g \int^t A_z },
    \label{eq:qsol_flux_fermi}
\end{equation}
with $k_j \equiv 2 \pi ( j + 1/ 2 )/L$.
We suppressed the index $f = 1 ,\cdots, N$ for brevity.
The creation/annihilation operators satisfy the usual canonical commutation relation of $\{\hat a_j, \hat a_k^\dag \} = \delta_{jk}$.
They transform as $\hat a_j^{R/L} \to \hat a_{j\mp 1}^{R/L}$ under the large gauge transformation of $\psi \to e^{2 \pi i z / L} \psi$ and $g A_z \to g A_z - 2 \pi / L$.

Now one may derive the quantum mechanics of $A_z$ by integrating out fermions, \textit{i.e.}, inserting the solution \eqref{eq:qsol_flux_fermi} into the fermion current.
For this purpose, we need to take an appropriate regularization, consistent with the gauge invariance.
The Hamiltonian of the fermions reads
\begin{align}
  \hat H_F (t) 
&= \oint \dd z \, \sum_{f=1}^N \left[ - i \psi_{R,f}^\dagger (\partial_z + i g A_z) \psi_{R,f} + i \psi_{L,f}^\dagger (\partial_z + i g A_z) \psi_{L,f}  \right] \\
&= \frac{1}{2} \sum_{f = 1}^N \sum_{j \in \mathbb{Z}} 
  \qty[ \omega_j^R (t) \qty( \hat a_j^{R \dag} \hat a_j^{R} - \hat a_j^{R} \hat a_j^{R \dag} )
  + \omega_j^L (t) \qty( \hat a_j^{L \dag} \hat a_j^{L} - \hat a_j^{L} \hat a_j^{L \dag} ) ],   \label{eq:H_flux_fermi}
\end{align}
where $\omega^{R/L}_j (t) \equiv k_j \pm g A_z (t)$.
We have dropped irrelevant constant terms in the second line.
The particles/anti-particles are defined as positive/negative frequency modes with $\omega^{R/L}_j \gtrless 0$ respectively.
One may introduce $g L A_z (t) = 2 \pi n(t) + g L \tilde A_z (t)$ such that $|g L \tilde A_z (t)| < \pi$,
and hence the positive/negative frequency modes correspond to $j \gtrless \mp n(t)$.
This Hamiltonian is invariant under the large gauge transformation as it should be.
Hence, we can make sense of the operator $\oint \hat J_z = \hat Q_5 = \hat Q_R - \hat Q_L$ with a regulator defined by the spectrum of $\hat H_F$:
\begin{align}
    \hat Q_{R/L} (t) \equiv \sum_{f = 1}^N \oint \dd z  ~\hat \psi^\dag_{R/L} \hat \psi_{R/L}
    &=  \frac{1}{2} \lim_{s \to 0} \sum_{f = 1}^N \sum_{j \in \mathbb{Z}} \qty( \hat a_j^{R/L \dag} \hat a_j^{R/L} - \hat a_j^{R/L} \hat a_j^{R/L \dag} ) R_s( \omega_j^{R/L}) ,
\end{align}
where $R_s (x)$ is the zeta-function regulator $R_s(x) \equiv |x|^{-s}$ with $\mathrm{Re}\, s>0$.
This can be further rewritten as 
\begin{align}
    \hat Q_{R/L} (t) 
    &=  \sum_{f = 1}^N \hat N_f^{R/L}[A_z(t)] + \hat Q_{R/L}^\text{vac}(t)
  \label{eq:q5_N+vac},
\end{align}
with
\begin{align}
  \hat N_f^{R/L}[A_z(t)] 
%    &\equiv  \sum_{j \in \mathbb{Z}} 
% \qty( \theta \left(\omega_j^{R/L}\right) \hat a_j^{R/L \dag} \hat a_j^{R/L} - \theta \left(-\omega_j^{R/L} \right) \hat a_j^{R/L \dag} \hat a_j^{R/L}) \\
    &\equiv \sum_{j \geq \mp n } \hat a_j^{R/L \dag} \hat a_j^{R/L} - \sum_{j \leq \mp n -1 }  \hat a_j^{R/L \dag} \hat a_j^{R/L},
\end{align}
\begin{align}
\hat Q_{R/L}^\text{vac}(t) \equiv - \lim_{s \to 0} \frac{1}{2}\sum_{f = 1}^N \sum_{j \in \mathbb{Z}}  \mathrm{sgn}\left(\omega_j^{R/L}\right) R_s( \omega_j^{R/L}),
\end{align}
where $\hat N_f^{R/L}[A_z(t)] $ counts the number of particles minus anti-particles associated with the Hamiltonian of \eqref{eq:H_flux_fermi} and has integer eigenvalues.
On the other hand, $\hat Q_{R/L}^\text{vac}$ takes non-integer values,
\begin{align}
\hat Q_{R/L}^\text{vac}(t)
% - \lim_{\Lambda \to \infty} \frac{1}{2}\sum_{f = 1}^N \sum_{j \in \mathbb{Z}}  \mathrm{sgn}\left(\omega_j^{R/L}\right) R( |\omega_j^{R/L}|/\Lambda) \\
%
=& - \lim_{s \to 0} \frac{1}{2}\sum_{f = 1}^N \qty (\sum_{j \geq \mp n } R_s( \omega_j^{R/L}) - \sum_{j \leq \mp n -1} R_s( \omega_j^{R/L} )) \\
%
%=& - \lim_{\Lambda \to \infty} \frac{1}{2}\sum_{f = 1}^N \qty (\sum_{j \geq 0} R( \frac{2\pi}{L} \qty(j+ \frac{1}{2} + x )) - \sum_{j \leq -1} R( \frac{2\pi}{L} \qty(j+ \frac{1}{2} + x )) ) \\
%
%=& - \lim_{\Lambda \to \infty} \frac{1}{2}\sum_{f = 1}^N \qty (\sum_{j \geq 0} R( \frac{2\pi}{L} \qty(j+ \frac{1}{2} + x )) - \sum_{j \leq -1} R( \frac{2\pi}{L} \qty(j+ \frac{1}{2} + x )) ) \\
%
=& \pm N \frac{gL \tilde A_z(t)}{2 \pi},
\end{align}
where we have used $\sum_{m=0}^\infty |m+\alpha|^{-s} \to 1/2 -\alpha $ as $s \to 0$.
Thus $\hat Q_{R/L}^\text{vac}(t)$ cannot be regarded as the number of the fermion particles, 
but as the contribution from the vacuum depending on the gauge zeromode, which is the Atiyah-Patodi-Singer $\eta$-invariant~\cite{Atiyah:1975jf} (See also \cite{Kobayashi:2021jbn}).\footnote{
Such non-trivial vacuum contribution becomes also important in the background of chromonatural inflation~\cite{Domcke:2018gfr}.
}
For later convenience, we also write down the following expression
\begin{equation}
  \hat Q_{R/L} (t) = \sum_f\hat N_f^{R/L} [0]
  \pm N \frac{g L  A_z (t)}{2 \pi},
  \label{eq:q5_qsol_flux}
\end{equation}
where the gauge-field dependent part of $\hat N_f^{R/L}$ is absorbed in $A_z$.

Eventually we obtain a one-dimensional effective theory, \textit{i.e.}, quantum mechanics, by inserting the solution \eqref{eq:q5_qsol_flux} into the action. 
The equation of motion reads
\begin{equation}
  0 = \frac{1}{g B / 2 \pi} \ddot{\hat{A}}_z
  - \frac{g \dot \theta}{2 \pi} 
  + g \qty[ \frac{\hat N _5[0]}{NL} + 2\frac{g \hat A_z (t)}{2 \pi} ]
  ~~\longrightarrow~~
  0 = \ddot{\hat{Q}}_5
  + \omega_B^2 \qty(  \hat Q_5 - 2N \frac{\dot \theta/2}{2 \pi} L ).
\end{equation}
with $\hat N _5[0] \equiv \sum_f \hat N_f^{R} [0] - \sum_f \hat N_f^{L} [0]$.
The right equation is obtained 
by noting that $\hat N_5 [0]$ is a time-independent operator.
It is solved by
\begin{align}
  \label{eq:gauge_zero_mode}
  \hat Q_5 (t) 
  = 2 NL \frac{g \hat A_z (t)}{2 \pi} + \hat N_5 [0] 
  &= Q_5 (t) + 
  \sqrt{ \frac{NL \omega_B}{2 \pi} } \qty( i \hat b_0 e^{- i \omega_B t} + \text{H.c.} ) \\
  &= 2N \times \frac{v_\theta/2}{2 \pi} L + 
  \sqrt{ \frac{NL \omega_B}{2 \pi} } \qty( i \hat B_0 e^{- i \omega_B t} + \text{H.c.} ),
\end{align}
where a $c$-number function $Q_5$ is given by
\begin{equation}
  Q_5 (t) = \begin{cases}
    0 &\text{for}~~~ t \leqslant 0, \\
    2 N \times \frac{v_\theta/2}{2 \pi} L \times \qty( 1 - \cos \omega_B t ) & \text{for}~~~ 0 < t .
    %\leqslant t_f \\ 2 N \times \frac{v_\theta/2}{2\pi} L \times \qty[ \cos \omega_B \qty( t - t_f ) - \cos \omega_B t ] & \text{for}~~~ t_f < t.
\end{cases}
\end{equation}
Recalling the rule of bosonization, \textit{i.e.}, $\hat Q_5 = N L \dot{\hat{\phi}}_0 / (2 \pi)$, 
we immediately reproduce the bosonized solution given in Eqs.~\eqref{eq:qsol_flux_M} or \eqref{eq:qsol_flux_b} up to an unphysical constant.

\paragraph{Interpretation.}
This result indicates that the boson theory automatically involves the both contribution of $\hat N_5 [0]$ and $\hat A_z$ in a particular combination so that its derivative, 
such as $\dot{\hat{\phi}}_0$ and $\partial_z \hat{\phi}_0$, is invariant under the large gauge transformation.
If we fix the gauge so that the gauge-field zero mode $A_z (t)$ always resides in a fundamental region specified by $| g L A_z (t) | < \pi$,
a $2N$-unit integer part of $Q_5$ is regarded as the fermion contribution while the rest is coming from $\tilde A_z$.
In other words, a trivial Wilson loop of $A_z$ along the flux can be interpreted as fermions while a remaining non-interger part of the chiral charge is carried by a non-trivial Wilson loop.
Hence, we should not be surprised by a non-integer chiral charge of $\langle \hat Q_5 \rangle$.

Still, the observation of individual fermions is not completely clear within this two-dimensional theory as the bosonized theory never distinguishes $\hat N_5 [0]$ from $\hat A_z$.
This is basically because its separation is gauge dependent within the two-dimensional theory. We need to recover the information of a higher dimension to some extent.
This issue would be resolved if we introduce another field $X$ that can propagate freely even in other ($x$ or $y$) direction, contrary to $\psi$, while extracting the chiral charge.
Assuming that the flux loop is confined in a finite box, we may define the observation unambiguously by detecting the $X$-particles outside the box. 
This is a higher dimensional analogue of the monopole case, 
\textit{i.e.}, the chiral charge is broken on the flux loop, while preserved in the bulk.
Though such an extension is beyond the scope of this paper, we expect that the chiral charge would be observed in $2 N$-unit through the $X$-particles, provided the monopole example.
See also discussion in Sec.~\ref{sec:concl}.

\section{Conclusions and discussion}
\label{sec:concl}

In this paper, we study the production of chiral charges by a time-dependent $\theta$ term in the background of the static magnetic field from a monopole or a flux loop.
We restrict our attention to the s-wave configuration for the monopole and the lowest Landau level for the magnetic flux loop since these modes are responsible for the chiral anomaly.
This simplification allows us to solve this system analytically by means of bosonization.

We find by this study that the chiral charge generation occurs in a somewhat interesting manner. 
For the case of the magnetic monopole, the flow of the charge by the time-dependent $\theta$ is cancelled by the emission of the opposite charge from the monopole so that the Witten effect is absent. 
The emission from the monopole is inevitably accompanied by the flow of the chiral charge that produce the chiral charge in the system.
For the case of the magnetic flux loop, the flow of the charge induces the electric field along the flux. 
This shifts the Fermi momentum of the left- and right-handed fermions asymmetrically, resulting in the chiral asymmetry.
In contrast to the magnetic monopole, the chiral symmetry is broken uniformly on the flux and hence the backreaction of produced fermions cannot be neglected.
As a result, the chiral charge eventually oscillates around some value indicated by the effective chemical potential for the chiral charge, \textit{i.e.}, $\dot \theta / 2$.

In both cases, we find that the expectation value of the chiral charge can be non-integer.
Moreover, in the case of the monopole, a non-integer electric charge of $\theta / (2 \pi)$ is emitted in order to cancel out the flow of the electric charge onto the monopole induced by the Chern-Simons coupling.
These results are puzzling at a first glance as the number of fermion particles is countable, which presumably is responsible for these charges.
In the case of the monopole, there is an unambiguous relation between the boson and fermion states in the bulk, 
which indicates that the resultant wave function is understood as a coherent summation over entangled pairs of fermion particle and its anti-particle.
Hence, it is not surprising that the expectation values of the chiral charge or the current exhibit a non-integer value.
In the case of the flux loop, on the other hand, a non-trivial Wilson loop of the gauge field $A_z$ along the flux can carry the chiral charge.
Again, the chiral charge can be non-integer because the expectation value of $\hat Q_5$ automatically contains contributions both from the fermions and the Wilson loop of $A_z$.

The bosonized picture is much simpler than the original fermion theory.
As mentioned in the introduction, the boson, \textit{i.e.}, an axion like degree of freedom, tries to cancel out the effect of the $\theta$ term.
However, this cancellation is not instantaneous. 
This dynamics of the axion, in the end, gives rise to the chiral asymmetry.
In the case of the monopole, this axion can exactly cancel out the $\theta$ term only at the location of the monopole, but its effect cannot propagate faster than the speed of light, 
which implies the production of the chiral charge and the electric current from the monopole.
In the case of the flux loop, 
the solution we obtain does not follow the motion of $\theta$ completely, indicating $E\cdot B$ generating.
In the contrary to the monopole case, this axion receives the negative backreaction, which results in an oscillating motion around a certain value.

Our discussion would imply some conditions for the baryon asymmetry generation by the slow rolling axion to be reliable.
In the cosmological applications, 
for instance, an axionic inflaton coupled to the $\theta$ term of $U(1)_Y$ leads to the simultaneous production of the baryon asymmetry and the hyper magnetic fields according to the SM chiral anomaly equation, which can be the origin of the current baryon asymmetry.
In this case, the hyper magnetic fields live in the flat four-dimensional spacetime, where the flux loop can be contracted.
Hence, the baryon charges have to be generated in a way so that the chiral fermions can be extracted from the flux.
Otherwise the baryon charge would vanish completely when the hyper magnetic fluxes collapse.
This consideration implies that the number of hyper magnetic fluxes inside a horizon should be larger than unity, and that the generated charges of each SM species should be large enough in a unit of $\Delta Q_{e} = 36, \Delta Q_{L} = -18, \Delta Q_u = 48, \Delta Q_d = 12, \Delta Q_{Q} = -6$ per a hyper magnetic flux for each generation of the right-handed electron, left-handed lepton, right-handed up-/down-type quarks, and left-handed quarks, respectively.
If these criteria are not met, the quantum fluctuations would be relevant as we have learned from the monopole example, and it is not clear how to match this state to the conventional picture of thermal particles propagating in the background gauge fields.
Note that our discussion is drawn from the result based on the effective two-dimensional setup. 
In order to understand the dynamics comprehensively, we need to describe the simultaneous production of chiral charges and magnetic links quantum mechanically, which requires the knowledge beyond the two dimensions.
This interesting question is left for future work.

\section*{Acknowledgement}
The authors would like to thank Satoshi Iso for useful discussions.
This work is supported by JSPS KAKENHI Grants No.~JP21J01117~(YH), No.~JP19H00689~(RK, RM), No.~JP21H01086~(RK) and MEXT KAKENHI Grant No.~JP18H05542~(RK).
KM was supported in part by MEXT Leading Initiative for Excellent Young Researchers Grant No.~JPMXS0320200430.

%%%%%%%%%%%%%%%%%%%%%%%%%%
\appendix
\section{The s-wave component of the fermion field in the monopole background}
\label{sec:appendix_spherical}
The s-wave component (\ref{eq:swave_fermion}) are determined so that
\begin{align}
  &\text{For}~\lambda_\text{rot}(x;R) \quad \text{s.t.}~ \overline A(x) = \overline A(R^{-1}x) -g^{-1} \dd \lambda_\text{rot}(x;R),\notag\\
  &\psi(x) = e^{i\lambda_\text{rot}(x;R)}\Lambda[R]\psi(R^{-1}x),
  \label{lambdaR}
\end{align}
where $R$ is an arbitrary rotation matrix and $\Lambda[R]$ gives the corresponding Lorentz transformation.

In the following, we show that Eq.~(\ref{eq:swave_fermion}) satisfies this requirement by extending the gauge group to $SU(2)$ and then reducing it to the diagonal $U(1)$ subgroup.
The $SU(2)$ gauge field $\overline{\mathcal{A}}$ for the monopole can be written so that a spatial rotation can be compensated by a global gauge transformation as
\begin{align}
  \overline{\mathcal{A}} = -\frac1{2g}\varepsilon_{abc}\sigma^a\frac{x^b}{r^2}\dd x^c
  =-\frac1{2g}\hat\varphi\cdot\vec\sigma \dd\theta + \frac1{2g}\hat\theta\cdot\vec\sigma \sin\theta \dd\varphi.
\end{align}
By the gauge transformation with respect to
\begin{align}
  U_0(x) := \begin{pmatrix}
    \cos\frac\theta2 & \sin\frac\theta2e^{-i\varphi}\\
    -\sin\frac\theta2e^{i\varphi} & \cos\frac\theta2
  \end{pmatrix},
  \label{eq:transf_to_u1}
\end{align}
the $SU(2)$ gauge field reduces to the $U(1)$ gauge field as
\begin{align}
  U_0(x)\overline{\mathcal{A}}U_0^\dag(x) - ig^{-1}U_0(x)\dd U_0^\dag(x)= -\sigma^3 \overline A(x),
\end{align}
where we have used
\begin{align}
  &\quad U_0(x)\hat \theta\cdot \vec\sigma U_0^\dag(x) = \cos\varphi\sigma_1 + \sin\varphi\sigma_2,\quad
  U_0(x)\hat \varphi\cdot \vec\sigma U_0^\dag(x) = -\sin\varphi\sigma_1 + \cos\varphi\sigma_2\notag\\
  &-ig^{-1}U_0(x)\dd U_0^\dag(x) =\frac1{2g}(-\sin\varphi\sigma_1 + \cos\varphi\sigma_2)\dd\theta 
    -\frac1{2g}\sin\theta(\cos\varphi\sigma_1 + \sin\varphi\sigma_2) \dd\varphi
    -\frac1{2g}(-1+\cos\theta)\sigma_3 \dd\varphi
\end{align}
[Note that the sign in front of the coupling $g$ is determined from the covariant derivative $D := \dd +ig A$.]
%The full (diagonalizable) symmetric form of the gauge field including $t,r$ components is
%\begin{align}
%  \mathcal A = -\hat r\cdot\vec\sigma \left( A_t(t,r) dt + A_r(t,r) dr\right) + \overline{\mathcal A}.
%\end{align}
We can check that, for $\overline{\mathcal A}$, a spatial rotation can be compensated by the gauge transformation,
\begin{align}
  \overline{\mathcal A}(x) = U_\text{rot}[R]\overline{\mathcal A}(R^{-1}x)U_\text{rot}^\dag[R],
\end{align}
where $U_\text{rot}[R]\in SU(2)$ is defined so that 
\begin{align}
U_\text{rot}[R] \sigma^a U_\text{rot}^\dag[R] = \sigma^b {R_b}^a
\label{eq:rot}
\end{align}
for an arbitrary rotation matrix $R$.
The spherically symmetric form of the left-handed Weyl fermion field $\chi(x)$ is
\begin{align}
  {[\chi(x)]_{\alpha}}_k= \frac1{\sqrt{8\pi}r}\left({\varepsilon_\alpha}_k\, \chi^1(t,r) + i{[\hat r\cdot \vec\sigma \varepsilon]_{\alpha}}_k\, \chi^2(t,r)\right),
  \label{eq:fermion_spherical}
\end{align}
where
\begin{align}
  \varepsilon := \begin{pmatrix} 0 &1\\ -1 &0 \end{pmatrix}.
\end{align}
Under a Lorentz transformation $\Lambda_{2\times 2}$ and a gauge transformation $U_g$, it transforms as
\begin{align}
  \chi(x) \rightarrow \Lambda_{2\times 2}\chi(x) U_g^T(x).
\end{align}
The field (\ref{eq:fermion_spherical}) is spherically symmetric in the sense that
\begin{align}
  U_\text{rot}[R]\chi(R^{-1}x)U^T_\text{rot}[R] = \chi(x)
  \label{eq:fermion_symmetric}
\end{align}
because $U\varepsilon U^T = \varepsilon$ for an arbitrary $U\in SU(2)$, and Eq.~(\ref{eq:rot}).
The $U(1)$ form of the s-wave fermion (\ref{eq:swave_fermion}) is obtained by acting the gauge transformation (\ref{eq:transf_to_u1}) as $\chi(x)\to\chi(x) U_0^T(x)$, which corresponds to the positively charged Dirac fermion $\psi$ as
\begin{align}
  \chi(x) U_0^T(x) = \left(\varepsilon \psi_R^*, \psi_L\right),\quad \psi(x) =\begin{pmatrix} \psi_L(x)\\ \psi_R(x)\end{pmatrix}.
\end{align}
We can see that the Lorentz and the $U(1)$ gauge transformations act correctly as
\begin{align}
  \Lambda\chi U_0^T(x) e^{-i\sigma_3 \lambda(x)} = \left(\varepsilon \left(e^{i\lambda(x)}(\Lambda^{-1})^\dag\psi_R\right)^*, e^{i\lambda(x)}\Lambda\psi_L\right),
  \label{eq:transf_fermion}
\end{align}
because of $\Lambda\varepsilon \Lambda^T= \varepsilon $.
Under the gauge transformation with respect to $U_0(x)$, the fields changes as
\begin{align}
  & U_0\mathcal AU_0^\dag -ig^{-1}U_0 \dd U_0^\dag= -\sigma_3\left(A_t \dd t + A_r \dd r + \overline A\right) = -\sigma_3 A,\notag\\
  &\chi U_0^T = \frac1{\sqrt{4\pi}r}U_0^\dag\begin{pmatrix} 0 & \psi_{L,s} \\- \psi_{R,s}^* & 0 \end{pmatrix} \notag\\
  &\hphantom{\chi U_0^T} = \frac1{\sqrt{4\pi}r}
  \begin{pmatrix}
  \sin\frac\theta2e^{-i\varphi}\ \psi^*_{R,s} & \cos\frac\theta2\ \psi_{L,s} \\ -\cos\frac\theta2\ \psi_{R,s}^* & \sin\frac\theta2e^{i\varphi}\ \psi_{L,s} 
  \end{pmatrix},\notag\\
  &\psi_{L,s} := (\chi_j^1 + i \chi_j^2)/\sqrt{2} ,\quad \psi_{R,s}^* := (\chi_j^1-i\chi_j^2)/\sqrt2,
\end{align}
where we have used
\begin{align}
  U_0(x)\hat r\cdot \vec\sigma U_0^\dag(x) = \sigma_3.
  \label{eq:diag_sigma}
\end{align}
Thus we obtain the form (\ref{eq:swave_fermion}).
We can also show that this form actually satisfies the condition (\ref{lambdaR}) as follows.
Let us consider the gauge transformation with respect to
\begin{align}
  \tilde U(x;R) := U_0(x)U_\text{rot}[R] U_0^\dag(R^{-1}x).
  \label{def_tildeU}
\end{align}
This matrix is an element of the little group with respect to $\sigma^3$:
\begin{align}
  \tilde U^\dag(x;R) \sigma^3 \tilde U(x;R) &= U_0(R^{-1}x)U^\dag_{\text{rot}}[R]\hat r\cdot \vec\sigma U_{\text{rot}}[R]U^\dag_0(R^{-1}x)\notag\\
  &=U_0(R^{-1}x)(R^{-1}\hat r)\cdot \vec\sigma U^\dag_0(R^{-1}x)\notag\\
  &= \sigma^3,
\end{align}
where we use Eqs.~(\ref{eq:rot}) and (\ref{eq:diag_sigma}).
Because the little group is the diagonal $U(1)$ subgroup, there exists a function $\lambda_{\text{rot}}(x;R)$ such that
\begin{align}
  \tilde U(x;R) = \exp\left(-i\sigma^3\lambda_\text{rot}(x;R)\right).
  \label{little}
\end{align}
We see that this function $\lambda_{\text{rot}}(x;R)$ satisfies the first line of Eq.~(\ref{lambdaR}) from the following sequence of the gauge transformation:
\begin{align}
  -\sigma^3 \overline{A}(x)\ \xrightarrow[\hspace{3em}]{U_0^\dag(x)}\ \overline{\mathcal A}(x)\ \xrightarrow[\hspace{3em}]{U^\dag_\text{rot}[R]}\ \overline{\mathcal A}(R^{-1}x)\ \xrightarrow[\hspace{3em}]{U_0(R^{-1}x)}\ -\sigma^3 \overline{A}(R^{-1}x).
\end{align}
For the spherically symmetric fermion field, it is satisfied that
\begin{align}
  &U_\text{rot}[R]\chi(R^{-1}x)U_0^T(R^{-1}x)\tilde U^T(x) = \chi(x)U_0^T(x),\notag\\
  \Leftrightarrow\quad &\left(\left(\varepsilon U_{\text{rot}}[R]\psi_R(R^{-1}x)\right)^*,U_{\text{rot}}[R]\psi_L(R^{-1}x)\right)\tilde U^T(x) = \left(\varepsilon\psi_R^*(x),\psi_L(x)\right),
\end{align}
which is shown by using the definition (\ref{def_tildeU}) of $\tilde U(x;R)$ and Eq.~(\ref{eq:fermion_symmetric}).
This confirms the second line of the condition (\ref{lambdaR}) due to Eq.~(\ref{little}), Eq.~(\ref{eq:transf_fermion}) and
\begin{align}
  \Lambda[R] = \begin{pmatrix}
  U_\text{rot}[R]&0\\
  0&U_\text{rot}[R]
  \end{pmatrix}.
\end{align}

\section{The anomaly equation in the monopole case}
\label{sec:A}
In this appendix, we show Eq.~(\ref{eq:topo_dens}) to confirm the anomaly equation at the leading order in $g$.
In order to show the equation, it is sufficient to confirm
\begin{align}
  \int_{B^3_\varepsilon} \dd^3 x\, \frac{g^2}{4\pi^2}\bm E\cdot\bm B \to \frac{v_\theta}{4\pi}\quad \text{as }g\to0
\end{align}
for infinitesimally small 3-ball $B^3_\varepsilon$ around the monopole.

Now we introduce the dimensionless variable $\tilde r = v_\theta r$ and $\tilde t = v_\theta t$.
Because the Bessel function is expanded as
\begin{align}
  J_\alpha(x) = \frac1{\Gamma(\alpha+1)}\left(\frac x2\right)^\alpha (1 + \mathcal O(x)),
\end{align}
the scalar $\tilde \phi \equiv\phi-\theta$ is expanded around $\tilde r=0$ as
\begin{align}
  \tilde \phi (\tilde t,\tilde r) = (-1+\mathcal O(g^2)) \tilde r^{1/2 +\alpha} + \mathcal O(\tilde r^2,g^2)
\end{align}
where we used the fact that $\tilde \phi\to - \tilde r$ as $g^2\to 0$.
Thus we obtain
\begin{align}
  \frac{g^2}{4\pi^2}\bm E\cdot\bm B 
  &= - \frac{g^2\tilde\phi}{64\pi^4r^4}\\
  &= - \frac{g^2}{64\pi^4r^4}\left(-1+\mathcal O(g^2)\right)\tilde r^{1/2+\alpha} + \mathcal O(\tilde r^{-2},g^2),
\end{align}
where the term $\mathcal O(g^2)$ does not depend on $\tilde r$.
The integral over $B^3_\varepsilon$ is calculated as
\begin{align}
  \int_{B^3_\varepsilon} \dd^3 x\, \frac{g^2}{4\pi^2}\bm E\cdot\bm B
   &= - v_\theta \frac{g^2}{16\pi^3}\left(-1+\mathcal O(g^2)\right)\int_0^\varepsilon \dd\tilde r \tilde r^{-3/2+\alpha} + \mathcal O(\varepsilon,g^2)\notag\\
   &= - v_\theta \frac{g^2}{16\pi^3}\left(-1+\mathcal O(g^2)\right)\frac1{-1/2+\alpha}\varepsilon^{-1/2+\alpha} + \mathcal O(\varepsilon,g^2)\notag\\
   &\to \frac{v_\theta}{4\pi} \quad\text{as}~~~g\to 0.
\end{align}

\section{The quntum fermion theory in the monopole background}

\subsection{Hamiltonian diagonalization and currents}
\label{sec:appendix_monopole}
In this appendix, we solve the fermion theory in the monopole background, whose action is given as
\begin{align}
  S_f = \int \dd t\int_0^{2L}\dd r \qty[i\psi_L^\dag \qty(\partial_t-\partial_r)\psi_L + i\psi_R^\dag \qty(\partial_t+\partial_r)\psi_R],
\end{align}
where the fermion fields satisfy the boundary condition
\begin{align}
  e^{i\theta}\psi_R|_{r=0}=\psi_L|_{r=0} ,\quad
  e^{i\theta}\psi_R|_{r=L} = -\psi_L|_{r=L},\quad \theta(t) = v_\theta t.
\end{align}
The general solution of the Dirac equation can be written as
\begin{align}
  \hat\psi_{R/L}(t,r)=\frac1{\sqrt{4L}}\sum_{n\in\mathbb Z}\hat A_n e^{-i(k_n\pm v_\theta/2)(t\mp r)},\quad k_n\equiv \frac{(n+1/2)\pi}{2L}.
  \label{expansion2}
\end{align}
Note that this expansion is different from Eq.~(\ref{expansion}).
The canonical anti-commutation relation
\begin{align}
\qty{\hat\psi_{R/L}(t,r),\hat\psi^\dag_{R/L}(t,r')} = i\delta(r-r') 
\end{align}
implies that $\hat A_n$ fulfills 
\begin{align}
\qty{\hat A_n,\hat A_m^\dag}=\delta_{nm}.
\end{align}
The Hamiltonian is not diagonalized in this basis at any time as
\begin{align}
  &\hat H(t) =\int_0^{2L}\dd r\,{:}\qty(i\hat\psi_L^\dag\partial_r\hat\psi_L-i\hat\psi_R^\dag\partial_r\hat\psi_R){:}
  =\frac{v_\theta}{\pi}\sum_{n\in\text{odd}}\qty(\hat B_ne^{-i\bar k_n t} + \text{H.c.}) + \qty(\sum_{n\in\mathbb Z}k_n\hat A_n^\dag \hat A_n)_{\text{reg}},\notag\\
  &-in\hat B_n \equiv \sum_{m\in\mathbb Z}\hat A_{m}^\dag \hat A_{m+n},\quad \bar k_n \equiv \frac{n\pi}{2L},
\end{align}
where $(\bullet)_{\text{reg}}$ denotes some regularization.
The operator $\hat B_n$ here actually corresponds to that in Eq.~(\ref{eq:qsol_monopole_M}) in the bosonized theory.
We will show in the following that the Hamiltonian is diagonalized as
\begin{align}
  \hat H(t) = \qty(\sum_{n\in\mathbb Z}k_n\hat V(t)\hat A_n^\dag \hat A_n V^\dag(t))_{\text{reg}}+\text{const.},\quad
  \hat V(t)\equiv \exp\qty(\sum_{n=1}^\infty \qty(nX_n(t)\hat B_n^\dag - nX_n(t)^*\hat B_n)),
\end{align}
where we define
\begin{align}
  X_n(t) \equiv \begin{dcases}
  0 & \text{for }n\in\text{even},\\
  -\frac{2Lv_\theta}{\pi^2n^2}e^{i\bar k_nt} & \text{for }n\in\text{odd}.
  \end{dcases}
\end{align}
The operator $\hat B_n$ satisfies
\begin{align}
  [\hat B_n,\hat A_m] = -i\frac1n\hat A_{m+n},\quad [\hat B_n,\hat A^\dag_m] = i\frac1n\hat A^\dag_{m-n},
  \label{commutatorBA}
\end{align}
which is shown by using the definition of $\hat B_n$ and the anti-commutation relation of $\hat A_n$.
Using these properties,  we can show that
\begin{align}
  [\hat B_m,\sum_{n\in\mathbb Z}A_n^\dag A_n] = 0,\quad [\hat B_m,\sum_{n\in\mathbb Z}nA_n^\dag A_n] = m\hat B_m.
  \label{commutatorBAA}
\end{align}
Note that these equation does not depend on the regulator because the c-number part coming from regulator drops due to the commutator.
Using this we obtain
\begin{align}
\qty[\sum_{n=1}^\infty \qty(nX_n(t)\hat B_n^\dag-nX_n^*(t)\hat B_n),\sum_{l\in\mathbb Z}k_l\hat A_l^\dag\hat A_l]
=\frac{v_\theta}{\pi}\sum_{n\in\text{odd}}\qty(B_ne^{-i\bar k_nt}+\text{H.c.}).
\end{align}
We also need the commutation relation for $\hat B_n$,
\begin{align}
  [\hat B_n,\hat B_m^\dag] = \delta_{nm}/n,
  \label{commutatorBB}
\end{align}
which can be shown by following Ref.~\cite{Iso:1988zi}.
When $n\neq m$, we can show the equation just by using the definition of $\hat B_n$.
To show the equation for $n=m$, we introduce the state $\ket{0}$ that is defined as the state satisfying $\hat A_n\ket{0}=0$ for $n\geq0$ and $\hat A^\dag_n\ket{0}=0$ for $n<0$.
Then we can show $[\hat B_n,\hat B_n^\dag]\ket{0}=(1/n)\ket{0}$.
Since the operator $[\hat B_n,\hat B_n^\dag]$ commutes with $\hat\psi_{R/L}(t,x)$, it commutes with any operator, which implies it is a c-number, \textit{i.e.}, $[\hat B_n,\hat B_n^\dag]=1/n$.
Using the commutation relation, we obtain
\begin{align}
\qty[\sum_{m=1}^\infty \qty(X_m(t)\hat B_m^\dag-X_m^*(t)\hat B_m),\qty[\sum_{n=1}^\infty \qty(X_n(t)\hat B_n^\dag-X_n^*(t)\hat B_n),\sum_{l\in\mathbb Z}k_l\hat A_l^\dag\hat A_l]] = \text{const.}
\end{align}
Using these properties we obtain
\begin{align}
  V(t)\qty(\sum_{n\in\mathbb Z}k_n\hat A_n^\dag\hat A_n)_{\text{reg}}V^\dag(t)
  &=\qty(\sum_{n\in\mathbb Z}k_n\hat A_n^\dag\hat A_n)_{\text{reg}} + \qty[\sum_{n=1}^\infty \qty(X_n(t)\hat B_n^\dag-X_n^*(t)\hat B_n),\sum_{n\in\mathbb Z}k_n\hat A_n^\dag\hat A_n] + \text{const.}\notag\\
  &=\hat H(t) + \text{const}.
\end{align}
Thus the creation/annihilation operators at $t=t_0$ are $\hat V(t_0)\hat A_n\hat V^\dag(t_0)$ and its Hermitian conjugation.
The vacuum $\ket{\Omega(t_0)}$ at $t=t_0$ is characterized by
\begin{align}
  \hat V(t_0)\hat A_n \hat V^\dag(t_0)\ket{\Omega(t_0)}= \hat V(t_0)\hat A_{-n-1}^\dag \hat V^\dag(t_0)\ket{\Omega(t_0)}= 0,\quad\text{for }n\in\mathbb Z_{\geq0}
\end{align}
Therefore we obtain the relation to the creation/annihilation operator in Eq.~(\ref{expansion}) as
\begin{align}
  \hat a_n(t_0) = \hat V(t_0)\hat A_n \hat V^\dag(t_0).
  \label{relation_aA}
\end{align}
Actually we directly show this equation using
\begin{align}
  \hat V(t_0)\hat\psi_{R/L}(t,r)\hat V^\dag(t_0) =e^{\mp i\phi_{R/L}(t-t_0\mp r)\mp iv_\theta(t-t_0\mp r)/2}\hat\psi_{R/L}(t,r),
\end{align}
which is obtained by using Eq.~(\ref{commutatorBA}).

In Section \ref{sec:monopole}, we stated that we should respect the "charge conjugation" symmetry (\ref{charge_conjugate_def}) to fix the c-number part of the currents, and determined it at the specific time $t=t_0$ as in Eq.~(\ref{c_t0}).
Now let us determine the c-number part at an arbitrary time.
As we showed in Eq.~(\ref{c_t0}), the c-number part vanishes when we expand the currents around the vacuum at that time, \textit{i.e.}, we can write
\begin{align}
  {:}\hat\psi_{R/L}^\dag(t,r)\hat\psi_{R/L}(t,r){:}
  &=\frac1{4L}\sum_{n=1}^{\infty}\left(\sum_{m\in\mathbb Z}\hat a^\dag_{m}(t)\hat a_{m+n}(t) e^{-i\bar k_n(t\mp r)} + \text{H.c.}\right)
  + \lim_{\varepsilon\to0}\frac1{4L}\sum_{n\in\mathbb Z}e^{-\varepsilon|k_n|}\left(\hat a^\dag_n(t)\hat a_n(t) - 1/2\right).
\end{align}
Using Eq.~(\ref{relation_aA}), we obtain
\begin{align}
  {:}\hat\psi_{R/L}^\dag(t,r)\hat\psi_{R/L}(t,r){:}
  &=-\frac1{4L}\sum_{n=1}^{\infty}\left( in\hat V(t)\hat B_n \hat V^\dag(t) e^{-i\bar k_n(t\mp r)} + \text{H.c.}\right)
  + \lim_{\varepsilon\to0}\frac1{4L}\sum_{n\in\mathbb Z}e^{-\varepsilon|k_n|}\left(\hat V(t)\hat A^\dag_n\hat A_n\hat V^\dag(t) - 1/2\right).
\end{align}
We can eliminate $V(t)$ and $V^\dag(t)$ in the last term because of the first one in Eq.~(\ref{commutatorBAA}).
Using the action of $V(t)$ to $\hat B_n$,
\begin{align}
  \hat V(t)\hat B_n\hat V^\dag(t) = \hat B_n - X_n(t),
\end{align}
which is obtained using the commutation relation (\ref{commutatorBB}) of $B_n$, and
\begin{align}
  \frac1{4L}\sum_{n=1}^\infty \qty(inX_n(t)e^{-i\bar k_n(t\mp r)} + \text{H.c.}) = 
  \pm\frac{v_\theta}{\pi^2}\sum_{n\in\text{odd}} \frac1n \sin(\frac{n\pi r}{2L})= \pm\frac{v_\theta}{4\pi}\quad \text{for }0<r<2L,
\end{align}
we obtain the expression of the currents in the expansion (\ref{expansion2}) as
\begin{align}
  {:}\hat\psi_{R/L}^\dag(t,r)\hat\psi_{R/L}(t,r){:}
  &=-\frac1{4L}\sum_{n=1}^{\infty}\left( in\hat B_n e^{-i\bar k_n(t\mp r)} + \text{H.c.}\right)
 + \lim_{\varepsilon\to0}\frac1{4L}\sum_{n\in\mathbb Z}e^{-\varepsilon|k_n|}\left(\hat A^\dag_n\hat A_n - 1/2\right) \pm \frac{v_\theta}{4\pi}.
\end{align}
By comparing the charge density $\hat J_{\bm 2}^0 = {:}\hat\psi_R^\dag\hat\psi_R{:}+{:}\hat\psi_L^\dag\hat\psi_L{:}$ and the chiral density $\hat J_{\bm 2,5}^0 = {:}\hat\psi_R^\dag\hat\psi_R{:}-{:}\hat\psi_L^\dag\hat\psi_L{:}$ in both theories, we see that $\hat B_n$ in both theories are identified.
Note that the $r$-independent part of the charge density $\hat J_{\bm 2}^0$ does not appear in the bosonized theory because we only treat $\hat Q_V=0$ sector.
We can also obtain the expression in the expansion (\ref{expansion}) around the vacuum at $t=t_0$ as
\begin{align}
  &{:}\hat\psi_{R/L}^\dag(t,r)\hat\psi_{R/L}(t,r){:}\notag\\
  &=-\frac1{4L}\sum_{n=1}^{\infty}\left( in\hat V^\dag(t_0)\hat b_n(t_0)\hat V(t_0) e^{-i\bar k_n(t\mp r)} + \text{H.c.}\right)
+ \lim_{\varepsilon\to0}\frac1{4L}\sum_{n\in\mathbb Z}e^{-\varepsilon|k_n|}\left(\hat a^\dag_n(t_0)\hat a_n(t_0) - 1/2\right) \pm \frac{v_\theta}{4\pi}\notag\\
  &=-\frac1{4L}\sum_{n=1}^{\infty}\left( in\hat b_n(t_0) e^{-i\bar k_n(t\mp r)} + \text{H.c.}\right)
+ \lim_{\varepsilon\to0}\frac1{4L}\sum_{n\in\mathbb Z}e^{-\varepsilon|k_n|}\left(\hat a^\dag_n(t_0)\hat a_n(t_0) - 1/2\right) -\frac1{2\pi}\partial_r\phi_{R/L}(t-t_0\mp r),
\end{align}
where we define $-in\hat b_n(t_0) \equiv \sum_{m\in\mathbb Z}\hat a_m^\dag(t_0)\hat a_{m+n}(t_0)$, and $\phi_{R/L}$ is defined in Eq.~(\ref{eq:classical_with_tf}).
Thus we obtain the c-number part as $c_{R/L}(t,r;t_0)=-\partial_r\phi_{R/L}(t-t_0\mp r)/(2\pi)$.

\subsection{Correspondence between the fermion field and the boson field}
The fermion field and boson field are related through the exact operator equation as \cite{VonDelft:1998pk}
\begin{align}
  \hat \psi_{R/L}(t,r) = \frac1{\sqrt{4L}}{:}e^{\mp i\hat\phi_{R/L}(t,r)}{:}e^{-i\qty(2\hat Q_V-1)\pi(t\mp r)/(4L)}\hat F.
  \label{correspondence2}
\end{align}
The operator $\hat F$ is the so-called Klein factor \cite{VonDelft:1998pk} or the "vacuum changing operator" \cite{Iso:1988zi}, which is defined so that
$\hat F\hat A_n = \hat A_{n-1}\hat F$, $\hat F\hat A_n^\dag = \hat A_{n-1}^\dag \hat F$, and $\hat F\ket0=\hat A_{-1}\ket0$, which implies%
\footnote{
The action of $\hat F^\dag$ is determined as follows. Just by taking the Hermitian conjugate, we obtain $\hat F^\dag\hat A_n=\hat A_{n+1}\hat F^\dag$, and $\hat F^\dag\hat A_n^\dag = \hat A_{n+1}^\dag \hat F^\dag$.
We also obtain $\hat F^\dag\ket0 = A_0^\dag\ket{0}$, because $\bra0A_0F^\dag\ket0 = \bra0F^\dag A_{-1}\ket0=\bra0A_{-1}^\dag A_{-1}\ket0 = 1$ and any other Fock state is orthogonal to $\hat F^\dag\ket0$.
Then we find that $\hat F\hat F^\dag = \hat F^\dag\hat F=1$, because $\hat F\hat F^\dag$ and $\hat F^\dag\hat F$ commute with $\hat A_n$ and $\hat A_n^\dag$ for any $n$ and $\hat F\hat F^\dag\ket0 = \hat F^\dag\hat F\ket0=\ket0$.
}
$\hat F\hat F^\dag =1$.
The left- and right-moving scalars are defined as
\begin{align}
  \hat \phi_{R/L} := \pm\sum_{n=1}^\infty\qty(\hat B_ne^{-i\bar k_n(t\mp r)} + \text{H.c.}) + v_\theta (t\mp r)/2,
\end{align}
which satisfies $\hat\phi_L + \hat \phi_R = \hat\phi$, the expansion of which is written in Eq.~(\ref{eq:qsol_monopole_M}).
The normal ordering ${:}\bullet{:}$ for the boson field is defined so that $\hat B_n$ is placed at the right end, \textit{i.e.},
\begin{align}
{:}e^{\mp i\tilde\phi_{R/L}(t,r)}{:}=\exp\qty(-i\sum_{n=1}^\infty\qty(\hat B_n^\dag e^{i\bar k_n(t\mp r)}))\exp\qty(-i\sum_{n=1}^\infty\qty(\hat B_ne^{-i\bar k_n(t\mp r)}))e^{\mp iv_\theta(t\mp r)/2}
\end{align}
Note that this is different from the normal ordering for the fermion field.
The charge $\hat Q_V$ can be written as
\begin{align}
  \hat Q_V = \lim_{\varepsilon\to0}\sum_{n\in\mathbb Z}e^{-\varepsilon|k_n|}(\hat A_n^\dag \hat A_n - 1/2),
\end{align}
which has integer eigenvalues.

The correspondence is shown as follows.
Any state in the Hilbert space can be expressed as
\begin{align}
  f\qty(\qty{\hat B_n^\dag})\ket{N},
\end{align}
where $f$ is an arbitrary function and $\ket N$ is "$N$-vacuum" defined so that $\hat A_{N+n}\ket N=\hat A_{N-n-1}^\dag\ket N=0$ for $n\geq0$.
Therefore, it is sufficient to show that the commutator with $B_n^\dag$ and the action on $\ket{N}$ is the same between the both side of Eq.~(\ref{correspondence2}).
Firstly we consider the commutator with $B_n^\dag$.
For the left hand side, using Eq.~(\ref{commutatorBA}), we show that
\begin{align}
  [\hat B_m^\dag, \hat\psi_{R/L}(t,r)] = i\frac1me^{-i\bar k_m(t\mp r)}\hat\psi_{R/L}(t,r).
\end{align}
Let us consider the right hand side.
From the definition of $\hat F$, we see that $[\hat B_n^\dag, \hat F]=0$.
Using Eq.~(\ref{commutatorBAA}), we obtain $[\hat B_n^\dag,e^{-i\qty(2\hat Q_V-1)\pi(t\mp r)/(4L)}]=0$.
By using Eq.~(\ref{commutatorBB}), we obtain
\begin{align}
  \qty[\hat B_m^\dag,\qty(-i\sum_{n=1}^\infty\qty(\hat B_ne^{-i\bar k_n(t\mp r)}))^k] = ik\frac1me^{-i\bar k_m(t\mp r)}\qty(-i\sum_{n=1}^\infty\qty(\hat B_ne^{-i\bar k_n(t\mp r)}))^{k-1}
\end{align}
It follows that
\begin{align}
  \qty[\hat B_m^\dag,{:}e^{\mp i\hat\phi_{R/L}(t,r)}{:}] =i\frac1me^{-i\bar k_m(t\mp r)}{:}e^{\mp i\hat\phi_{R/L}(t,r)}{:}
\end{align}
Thus we show that the commutator with $\hat B_n^\dag$ is the same for the both side of Eq.~(\ref{correspondence2}).
Next we show the action on $\ket N$ is the same for the both side.
Because $[\hat B_m,\hat \psi_{R/L}] = -i\exp(\smash{i\bar k_n(t\mp r)})\hat\psi_{R/L}/m$, the state $\psi_{R/L}(t,r)\ket{N}$ is the eigenstate of $\hat B_n$ belonging to the eigenvalue $-i\exp(\smash{i\bar k_n(t\mp r )})/m$.
We can show in the same way that the action of the right hand side on $\ket{N}$ is the eigenstate belonging to the same eigenvalue.
Hence they are proportional to each other.
The overall factor is also the same because
\begin{align}
  \bra N \hat F^\dag \hat\psi_{R/L}(t,r)\ket N = \bra N \hat A_{N-1}^\dag \hat\psi_{R/L}(t,r)\ket N
  = \frac1{\sqrt{4L}}e^{-i(k_{N-1}\pm v_\theta/2)(t\mp r)},
\end{align}
and
\begin{align}
  \bra N \hat F^\dag\frac1{\sqrt{4L}}{:}e^{\mp i\hat\phi_{R/L}(t,r)}{:}e^{-i\qty(2\hat Q_V-1)\pi(t\mp r)/(4L)}\hat F \ket N
  =\frac1{\sqrt{4L}} e^{-i(2N-1)\pi(t\mp r)/(4L)\mp iv_\theta(t\mp r)/2}.
\end{align}
Thus we show Eq.~(\ref{correspondence2}).

In the bosonized theory, the fermion operator $\hat\psi_{R/L}(t,r)$ is regarded as a creation operator of a sharp anti-kink, which corresponds to the positively charged fermion.
To see this, we should redefine the boson field as
\begin{align}
  \hat\Phi(t,r) = \hat\phi(t,r) - \frac{2\pi\qty(\hat Q_v - 1/2)}{2L} r.
\end{align}
In this definition, the all exponent in the correspondence (\ref{correspondence2}) is absorbed in the boson field. 
For this field, we obtain the following commutation relation with the fermion field:
\begin{align}
  [\hat\Phi(t,r),\hat\psi_{R/L}(t,r')] = -\qty(\sum_{n=1}^\infty\frac4n\cos(\bar k_n r')\sin(\bar k_n r) + \frac{2\pi}{2L} r)\hat\psi_{R/L}(t,r') = -2\pi\Theta(r-r')\hat\psi_{R/L}(t,r'),
\end{align}
where $\Theta(r)$ is the step function.
The coherent state with respect to the configuration $\Phi_c(r)$ can be written as $\hat U_c\ket{\Omega(t)}$ using the unitary operator $\hat U_c$ satisfying $[\hat\Phi(t,r),\hat U_c] = \Phi_c(r)\hat U_c$.
Thus, we can regard that the operator $\hat U_c$ creates the configuration $\Phi_c(r)$.
By analogy with this, we can regard the operator $\hat\psi_{R/L}(t,r)$ as a creation operator of a sharp anti-kink.
Note that due to the discontinuity of the sharp kink, its creation operator cannot be unitary.

\subsection{Probability to observe fermions in the monopole background}
\label{sec:appendix_probability}
We give an explicit calculation of the probability to observe a two-particle state in the monopole background discussed in the last part of Section \ref{sec:monopole}.
Firstly we show that the overlap between the vacuum $\ket{\Omega(0)}$ and the two-particle state at $t=t_f$ is nonzero only when $t_f=2\pi/v_\theta$ in the $L\to\infty$ limit.
The probability is written as
\begin{align}
  P(t_f):=\sum_{n,m\in\mathbb Z}|\bra{\Omega(0)}\hat a_n^\dag(t_f)\hat a_m(t_f)\ket{\Omega(t_f)}|^2.
\end{align}
We can rewrite this using the fermion fields at $t=t_f$ as
\begin{align}
  P(t_f)=\int_0^{2L}\dd r \int_0^{2L}\dd r'\, \qty|\bra{\Omega(t_f)}\hat V(t_f)\hat V^\dag(0){:}\hat\psi^\dag(t_f,r)\hat\psi(t_f,r){:}\ket{\Omega(t_f)}|^2 ,
  \quad \hat\psi(t_f,r) := \hat\psi_L(t_f,r) + \hat\psi_R(t_f,r).
  \label{prob}
\end{align}
Here we have used
\begin{align}
  \int_0^{2L}\dd r \hat\psi^\dag(t_f,r)\ket{\Omega(t_f)}\bra{\Omega(t_f)}\hat\psi(t_f,r) = \sum_{n\in\mathbb Z}\hat a_n^\dag(t_f)\ket{\Omega(t_f)}\bra{\Omega(t_f)}\hat a_n(t_f),
\end{align}
which is the projector to the Hilbert space of the one-particle state at $t=t_f$. 
Thus the creation operator of the fermion at the position $r$ should be the sum of the left- and right-mover differently from the case in the compact space.
Using the correspondence (\ref{correspondence2}), we can rewrite the two-particle state using the boson field.
In doing so, we need to be careful with the difference of the normal ordering between the fermion and the boson.
Consequently we find that
\begin{align}
  {:}\hat\psi_I^\dag(t_f,r)\hat\psi_J(t_f,r'){:} &= {:}e^{\eta_Ii\hat\phi_I(t_f,r)-\eta_Ji\hat\phi_J(t_f,r')}{:}D_{IJ}(r,r') - D_{IJ}(r,r'),
\end{align}
where $\eta_{R/L}=\pm1$.
We obtain the large $L$ dependence of the integrand in Eq.~(\ref{prob}) using the commutator (\ref{commutatorBB}) of $\hat B_n$ as
\begin{align}
 \qty|\bra{\Omega(t_f)}\hat V(t_f)\hat V^\dag(0){:}\hat\psi^\dag(t_f,r)\hat\psi(t_f,r){:}\ket{\Omega(t_f)}|^2  \sim \qty(\frac1{v_\theta L})^{\qty(\frac{v_\theta t_f}{2\pi})^2 - 2\frac{v_\theta t_f}{2\pi} + 1},
\end{align}
which approaches zero when $v_\theta t_f \neq 2\pi$.
We have numerically confirmed that the integrand in Eq.~(\ref{prob}) is localized in the region near $r=t$, and the integral does not give additional $L$ dependence. Thus the probability $P(t_f)$ is zero when $v_\theta t_f\neq 2\pi$.
We also calculate the full probability numerically, and obtain the value around $0.86$ when $v_\theta t_f =2\pi$.

As discussed in the last part of Section \ref{sec:monopole}, $P(t_f)=0$ does not mean that we cannot observe fermions.
An example of states that can be regarded as a two-particle state and have finite overlap with $\ket{\Omega(0)}$ is
\begin{align}
  {:}\hat\psi^\dag(t_f,r)\hat\psi(t_f,r'){:}\ket{\Omega(2\pi/v_\theta)},
\end{align}
which is interpreted as the state obtained by creating two-particles at $t=t_f$ in the state that will be the vacuum at $t=2\pi/v_\theta$.
The configuration of this state corresponds to the right panel of Fig. \ref{fig_phi_prof}.

%%%%%%%%%%%%%%%%%%%%%%%%%%%%%%%%%%%%%%%%%%%%%%%%%%%%%%%%%%%%%%%%%%%%%%%%%%%%%%%%%%%%%%%%%%%%%%%%%%%%

\newpage
%%%%%%%%%%%%
\small
\bibliographystyle{JHEP}
\bibliography{refs}
%%%%%%%%%%%%

\end{document}